\documentclass[letterpaper, 10 pt, conference]{ieeeconf}  

\IEEEoverridecommandlockouts              

\usepackage{cite}
\usepackage{amsmath,amssymb,amsfonts}
\usepackage{graphicx}
\usepackage{textcomp}
\usepackage[dvipsnames]{xcolor}
\def\BibTeX{{\rm B\kern-.05em{\sc i\kern-.025em b}\kern-.08em
    T\kern-.1667em\lower.7ex\hbox{E}\kern-.125emX}}

\usepackage{comment}

\usepackage[normalem]{ulem}

\usepackage{mleftright}

\usepackage{import}

\usepackage{algorithm}
\usepackage{algpseudocode}

\usepackage{pgfplots}
  \usepgfplotslibrary{groupplots}
  \usepgfplotslibrary{fillbetween} 
\pgfplotsset{compat=newest}

\usepackage{subcaption}

\usepackage{tikz}
\usetikzlibrary{positioning, arrows, calc, shadings}
\usetikzlibrary{fit}

\tikzstyle{blockdiag}	= [node distance=5mm, >=stealth', semithick]
\tikzstyle{block}			= [draw, rectangle, minimum width=1cm, minimum 
height=.8cm]
\tikzstyle{sum} = [draw,circle,inner sep=0pt, minimum size=6pt]
\tikzstyle{connector} = [draw,circle,inner sep=0pt, minimum size=2pt, 
fill=black]

\usepackage{array}
\newcommand{\vspacelem}{\vspace{0.1in}}

\newcommand{\norm}[1]{\left\|#1\right\|}
\newcommand{\abs}[1]{\left|#1\right|}
\newcommand{\field}[1]{\mathbb{#1}}
\newcommand{\R}{\field{R}}

\newcommand{\Sy}{\field{S}}
\newcommand{\N}{\field{N}}
\newcommand{\D}{\mathbb{D}}
\newcommand{\DHyp}{\mathbb{DH}}

\newcommand{\Msec}{\mcl{M}_\text{sec}}
\newcommand{\Mslope}{\mcl{M}_\text{slp}}
\newcommand{\Mislope}{\mcl{M}_\text{slp}^\text{inc}}
\newcommand{\Mcrelu}{\mcl{M}_\text{cReLU}}
\newcommand{\Micrelu}{\mcl{M}_\text{cReLU}^\text{inc}}
\newcommand{\Mbbnd}{\mcl{M}_\text{bnd}^\text{b}}
\newcommand{\Mbhsec}{\mcl{M}_\text{hsec}^\text{b}}
\newcommand{\MblReLU}{\mcl{M}_\text{ReLU}^\text{b}}
\newcommand{\M}{\mcl{M}}

\newcommand{\MbReLUc}{\mcl{M}_\text{ReLU,c}^\text{b}}
\newcommand{\MblReLUCOP}{\mcl{M}_\text{ReLU}^\text{b,COP}}
\newcommand{\MblReLUNN}{\mcl{M}_\text{ReLU}^\text{b,NN}}

\newcommand{\COP}{\mathbb{COP}}

\newcommand{\PSD}{\mathbb{PSD}}

\newcommand{\bmtx}{\begin{bmatrix}}
\newcommand{\emtx}{\end{bmatrix}}
\newcommand{\bsmtx}{\left[ \begin{smallmatrix}} 
\newcommand{\esmtx}{\end{smallmatrix} \right]} 
\newcommand{\bmatarray}[1]{\left[\begin{array}{#1}}
\newcommand{\ematarray}{\end{array}\right]} 

\newcommand{\mcl}[1]{\mathcal{#1}}

\definecolor{blue1}{RGB}{222,235,247}
\definecolor{blue2}{RGB}{158,202,225}
\definecolor{blue3}{RGB}{49,130,189}

\newcommand{\StateWrap}[1]{%
  \State \parbox[t]{\dimexpr\linewidth-\algorithmicindent\relax}{#1}}

\newcommand{\StateTwo}[2]{%
    \State
    \begin{tabular}[t]{@{}l@{}}
        #1\\
        #2
    \end{tabular}%
}
\newtheorem{mytheo}{Theorem}

\newtheorem{defn}{Definition}

\newtheorem{lem}{Lemma}
\newtheorem{prop}{Proposition}

\newcommand{\edtn}[1]{\null} 

\title{\LARGE \bf Lipschitz Bounds for Neural Networks Using a Two-Copy Linear Fractional Transformation
}

\author{Felix Biert\"umpfel, Bin Hu, and Peter Seiler
\thanks{*This work was supported by the European Union under Grant No. 101153910 and by the AFOSR under Grant \#FA9550-23-1-0732. The responsibility for the content of this paper is with its authors.}
\thanks{F. Biert\"umpfel is with the Department of Aerospace Engineering at Auburn University, Auburn, {\tt\small feb0013@auburn.edu}.}%
\thanks{B. Hu is with the Department of Electrical and Computer 
Engineering at the University of Illinois at Urbana-Champaign 
({\tt \small binhu7@illinois.edu}).}
\thanks{P. Seiler is with the Department of Electrical and Computer Engineering at the University of Michigan, Ann Arbor, {\tt\small pseiler@umich.edu}.}%
}

\begin{document}

\maketitle
\thispagestyle{empty}
\pagestyle{empty}

\begin{abstract}

This paper proposes a new method to bound the global Lipschitz constant of a neural network (NN).  
Our formulation uses a two-copy linear fractional transformation (LFT) representation of the NN with stacked signals.
We use biased quadratic constraints (QCs) to bound the input/output behavior of the nonlinear activation functions.
We then formulate a linear matrix inequality condition, using these biased QCs, to bound the Lipschitz constant of the NN.
The best (smallest) Lipschitz bound is computed via a convex, semidefinite program.  There are two key 
distinctions of the proposed two-copy formulation in comparison to existing methods that rely on a single incremental 
representation with incremental signals. First, the NN bias terms are eliminated in the standard incremental 
representation, but appear in our formulation. Second, there is a larger class of QCs that can be expressed 
for the nonlinearities in our formulation than in the incremental representation.
We show that our proposed formulation yields less conservative Lipschitz bounds due to these two differences.
We illustrate these points via several simple, motivating examples. We also demonstrate the proposed method using
a benchmark problem with a NN controller for an inverted pendulum.
\end{abstract}


\section{Introduction}\label{sec:intro}

This paper proposes a new method to bound the global Lipschitz constant of a neural network (NN). Such bounds are useful for certifying robustness properties of neural networks (NNs).  A variety of methods have been proposed for computing Lipschitz bounds and here we focus on SDP-based methods. Existing approaches, e.g., LipSDP~\cite{fazlyab2019efficient}, rely on an incremental formulation for the NN.  The input/output behavior of the nonlinear activation functions are characterized by quadratic constraints (QCs) or integral quadratic constraints (IQCs)~\cite{megretski97system,veenman2016,hu2017}. These constraints are then incorporated into a linear matrix inequality (LMI) condition that yields a global Lipschitz bound. The best (smallest) bound can be computed via a semidefinite program (SDP). Extensions address non-slope-restricted activations~\cite{Pauli2024} and local Lipschitz and other incremental input/output properties~\cite{Hashemi2021}. Related QC- and IQC-based methods have been developed for stability analysis of neural-network feedback systems~\cite{Yin2022,Pauli2021ZF} and for constructing recurrent neural networks with certified properties~\cite{Revay2021}.  A recent overivew is given in ~\cite{Manchester2026}.

These incremental formulations have two  limitations. First, the network biases do not appear in the incremental formulation. Thus, the same bound is obtained for all bias choices associated with fixed weights. Second, QCs must be formulated solely in terms of increments of the nonlinearity inputs and outputs. It is not possible to capture all joint constraints that exist on two separate copies of the nonlinearity.  The incremental form thus limits the class of QCs that can be
exploited.   We show that either limitation can make the resulting incremental gain bound arbitrarily conservative.  Similar issues have been observed in the literature. For example,~\cite{gronqvist22} studies the effect of neglected biases and shows that incremental-QC bounds can grow exponentially with repeated network layers. Related observations appear in~\cite{pauli21training,Gramlich2026}. More recently,~\cite{Kuang2026} presented examples for which common SDP-based verification methods fail, including ReLU networks with dead neurons. These observations motivate a unified formulation that retains bias information and permits richer QCs on the activation functions.

We address these limitations using a two-copy linear fractional transformation (LFT) representation of the NN with stacked signals. The two copies of the system are driven by independent inputs, but share the same constant input tied to the biases. This representation also enables biased, non-incremental QCs that jointly constrain the two copies of the nonlinearity.  The formulation yields an LMI condition and SDPs can then be used to compute the smallest Lipchitz bound within our two-copy LFT formulation.
Related QC-based formulations retain bias information for a single 
copy of the NN. In particular, \cite{fazlyab2020safety} considers safety and robustness analysis over prescribed input uncertainty sets, while \cite{Ebihara2024local} considers local Lipschitz analysis around a target input. In contrast, our formulation
simultaneously represents two copies of the NN corresponding to two
independent inputs and applies biased QCs directly to their stacked
nonlinear signals. This allows the bias information to be preserved
in a global incremental analysis and avoids restricting the nonlinear
description to QCs expressed only in terms of signal differences.
Here, we specialize this two copy representation to bound the global Lipschitz constant of feedforward NNs. The construction also extends to other incremental analyses, e.g. incremental $\ell_2$ bounds for dynamical systems.

Our paper has three main contributions. First, in Section~\ref{sec:motivation} we formally characterize two sources of conservatism inherent in the standard incremental formulation: the elimination of bias information and the restriction of QCs formulated solely on incremental signals. We show that these restrictions can lead to arbitrarily conservative incremental gain bounds. Second, we introduce a two-copy LFT representation of the NN with stacked signals. This representation retains the bias information and enables the use of richer, non-incremental biased QCs (Section~\ref{sec:main}) by using independent input signals. Third, for ReLU nonlinearities, we derive a complete characterization of biased QCs based on copositivity (Section~\ref{sec:main}). This extends the results in~\cite{Noori2025} to the biased setting. 
To enable the computational use of this complete characterization,
we design a tailored cutting-plane algorithm (Section~\ref{sec:comp}). We demonstrate the proposed approach on multiple examples in Section~\ref{sec:numex} including a NN with a dead neuron, a NN with exponentially growing Lipschitz bound, and  a simple NN-based controller of an inverted pendulum.

\vspacelem
\paragraph*{Notation}
We denote the set of $n$-dimensional real column vectors by $\R^n$. The set of vectors with non-negative entries, i.e., the nonnegative orthant, is denoted by $\R^{n}_{\ge 0}$. The standard $i^\text{th}$ basis vector of an $n$-dimensional space is denoted $e_i\in \R^{n}$. This vector has a $1$ entry at position $i$, while all other entries are zero. A column vector of dimension $n$ with all ones is denoted by~$\mathbf{1}$.

The set of $n\times m$ real matrices is denoted by $\R^{n\times m}$. The sets of $n \times n$ real, symmetric  and diagonal matrices are denoted by $\Sy^n$ and $\D^n$, respectively. Sets of matrices with nonnegative elements are denoted by $\R^{n\times m}_{\ge 0}$, $\Sy^n_{\ge0}$, and $\D_{\ge0}^n$, respectively. Similarly, $\D_{\pm 1}^n \subset \D^n$ denotes diagonal matrices whose diagonal terms are restricted to $\pm 1$. A matrix is doubly hyperdominant if it has non-positive off-diagonal elements with all row sums and column sums that are non-negative.
The set of $n\times n$ doubly hyperdominant matrices is denoted by $\DHyp^n$.  
We further denote the sets of positive semidefinite and copositive matrices as
\begin{align}\label{eq:psd}
    \PSD^n & := \{M\in \Sy^n:v^\top M v\ge 0 \, \, \forall v\in\R^n\}, \\
\label{eq:cop}
    \COP^n & := \{M\in \Sy^n:v^\top M v\ge 0 \, \, \forall v\in\R_{\ge0}^n\}.
\end{align}
The superscript is omitted when the dimension $n$ is clear from context.

The set $\N$ denotes the set of non-negative integers.  A real, vector-valued sequence $v$ is a function $v:\N \to \R^{n_v}$. The inner product of two such sequences $v$ and $w$ is defined by $\langle v, w \rangle := \sum_{k=0}^\infty v(k)^\top w(k)$. The set $\ell_2$ consists of sequences $v$ that satisfy $\langle v, v \rangle < \infty$. This is an inner product space and the corresponding norm is $\norm{v}_2 =\sqrt{\langle v,v\rangle}$.

\section{Problem Formulation}\label{sec:problem}

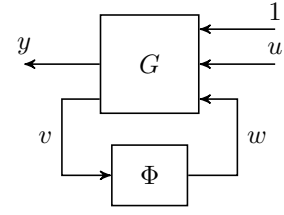
\begin{figure}[ht!]
\centering
\begin{tikzpicture}[blockdiag]
	\node[block,minimum width=1.3cm, minimum height=1.3cm](Plant) {$G$};
	\node[block, below= 0.4cm of Plant] (Delta) {$\Phi$};

	\draw[<-] ($(Plant.south east)!.5!(Plant.north east)$) -- +(+1.0cm, 0) 
	node[above, name=e]{$u$};

\draw[<-] ($(Plant.south east)!.85!(Plant.north east)$) -- +(+1.0cm, 0) 
	node[above, name=e]{$1$};

    \draw[->] ($(Plant.south west)!.5!(Plant.north west)$) -- +(-1.0cm, 0) 
	node[above, name=d]{$y$};
	
	\draw[<-] ($(Plant.south east)!.15!(Plant.north east)$) -- +(+0.5cm, 0) |- 
	node[near start, right]{$w$} (Delta);	
	\draw[->] ($(Plant.south west)!.15!(Plant.north west)$) -- +(-0.5cm, 0) |- 
	node[near start, left ]{$v$} (Delta);

\end{tikzpicture}
\caption{Feedback interconnection $F_\text{l}(G,\Phi)$ of a known matrix $G$ and nonlinearity $\Phi$.}
\label{fig:UncSys}
\vspace{0pt}
\end{figure}

We consider a neural network (NN) $f:\R^{n_u} \to \R^{n_y}$ mapping an input $u\in \R^{n_u}$ to an output $y\in\R^{n_y}$.  We represent the NN as the interconnection shown in Figure~\ref{fig:UncSys}. This interconnection consists of a nonlinearity~$\Phi:\R^m \to \R^m$ in feedback around a known $(n_y+m)\times(n_u+m+1)$, real matrix~$G$. This interconnection, denoted as $F_\text{l}(G,\Phi)$, is called a lower linear fractional transformation (LFT). 

Figure~\ref{fig:UncSys} corresponds to the following equations:
\begin{subequations}
\label{eq:LFT}
\begin{align}
\label{eq:LFTy}
        y &= G_{11} + G_{12}\,  u + G_{13}\, w, \\
\label{eq:LFTv}
        v &= G_{21} + G_{22}\,  u + G_{23}\,  w, \\
\label{eq:LFTw}
        w &= \Phi(v).
\end{align}
\end{subequations}
The vectors $G_{11} \in \R^{n_y}$ and $G_{21} \in \R^m$ represent bias terms in~\eqref{eq:LFTy} and~\eqref{eq:LFTv}. The function $\Phi$ is a static, repeated nonlinearity formed from $m$ copies of a scalar activation function $\phi:\R \to \R$. Specifically, the function $\Phi$ maps $v$ to~$w$ elementwise via $w_i = \phi(v_i)$ for~$i=1,\dots,m$. 
The lower LFT \eqref{eq:LFT} is in the form of an implicit (equilibrium) NN \cite{elghaoui21,bai19deep}.
This  models a wide range of common static NN architectures.

Note that plugging~$w = \Phi(v)$ into  \eqref{eq:LFTv} yields
\begin{equation}\label{eq:implicit}
    v = G_{21}  + G_{22} \, u + G_{23} \, \Phi(v).
\end{equation}
This an implicit equation that must be solved for $v$ as a function of $u$. The NN is said to be well-posed if there exists a unique solution~$v$ for any $u\in \R^{n_u}$.  If $G_{23}$ is strictly lower triangular, then \eqref{eq:implicit} can be solved for a unique $v$ and the NN is well-posed.\footnote{This follows because $\Phi$ is a repeated nonlinearity, $w_i=\phi(v_i)$ for $i=1,\ldots m$. If $G_{23}$ is strictly lower triangular then  the $i^{th}$ entry of $G_{23} \Phi(v)$ only depends on  $\{v_1,\ldots,v_{i-1}\}$. Hence we can use the first row of  \eqref{eq:implicit} to solve for $v_1$ and then work forward to solve for $\{v_2,\ldots,v_m\}$.} This structure appears in feedforward NNs as shown in the example below. In general, additional conditions are required to verify well-posedness. These can be derived from contraction mapping arguments (Theorem 2.2 in \cite{elghaoui21}) or using properties of convex optimizations (Theorem 1 and Proposition 2 in \cite{revay20lipschitz}).

Our goal is to verify that a well-posed NN, modeled by the lower LFT \eqref{eq:LFT}, has a Lipschitz constant bounded by $\gamma$. A  formal definition for the Lipschitz constant is provided next. 

\vspacelem
\begin{defn}[Lipschitz Constant]
A function $f:\R^{n_u} \to \R^{n_y}$ is Lipschitz continuous if there exists a finite constant $\gamma\ge 0$ such that $\|f(\bar{u}) - f(\hat{u})\|_2 \le \gamma \, \|\bar{u}- \hat{u}\|_2$ for all $\bar{u}$, $\hat{u}\in\R^{n_u}$. The smallest such $\gamma$ is called the Lipschitz constant of the function $f$.
\end{defn}
\vspacelem

Next, we show, for illustration, that a feedforward NN can be expressed as an lower LFT. Consider a multi-layer perceptron (MLP) $f:\R^{n_u} \rightarrow \R^{n_{y}}$ with ${L}$-layers.
The MLP is a feedfoward NN defined by the weights $W^l \in \R^{n_{l}\times n_{l-1}}$ and biases $b^l \in \R^{n_{l}}$ for $l=1,\ldots,L+1$. The NN maps $u$ to $y = f(u)$ (with $n_0:=n_u$ and $n_{L+1}:=n_y$) by 
\begin{equation}\label{eq:NNcomp}
\begin{split}
w^0 & = u \\
w^{l} &= \phi^l(W^lw^{l-1} + b^l)
       \,\, \text{ for }\, l = 1, \ldots, {L}\\
y&=W^{L+1} w^{L} + b^{L+1},
\end{split}
\end{equation}
The activation functions $\{ \phi^l\}_{l=1}^{L}$  are repeated nonlinearities obtained by applying the scalar function $\phi$ elementwise.

We can write the MLP as a lower LFT \eqref{eq:LFT}. First, define  $v^l := W^l w^{l-1} + b^l$ and let $v$ denote the vector obtained by stacking $v^1$, $v^2$, $\ldots$, $v^{L}$. The dimension of the vector $v$ is $m= \sum_{l=1}^{L} n_l$. Moreover, let $w$ denote the vector obtained by stacking $w^1$, $w^2$, $\ldots$, $w^L$ (but not including $w^0=u$). The dimension of $w$ is also $m$. The stacked vectors are related by $w=\Phi(v)$ where $\Phi:\R^m\to\R^m$ is a repeated nonlinearity with $m$ copies of the scalar activation function $\phi$. 

The MLP in \eqref{eq:NNcomp} is represented by the LFT $F_\text{l}(G,\Phi)$ with the repeated nonlinearity $\Phi$ and the matrix $G$ defined by
\begin{align*}
   G_{11} &= b^{L+1},
\,\,   
   G_{12} = 0_{n_y \times n_u}, \\
   G_{13} &= \bmtx 0_{n_y \times (m-n_{L})} & W^{L+1} \emtx, \\ 
     G_{21} &= \bmtx b^1 \\ b^2 \\ \vdots \\ b^{L}\emtx, 
\,\,
   G_{22} = \bmtx W^1 \\ 0_{n_2\times n_u} \\ \vdots \\ 0_{n_L\times n_u}\emtx,\\
G_{23} &= \bmtx 0_{n_1\times (m-n_{L})} & 0 \\ \text{blkdiag}(W^{2}, \ldots. W^{{L}}) & 0_{(m-n_{1})\times n_L} \emtx.
\end{align*} 
The matrix $G_{23}$ is strictly lower triangular. Hence, the MLP is well-posed, i.e. \eqref{eq:implicit} has a unique solution $v$ for any $u$. 

Note that the inputs of $G$ in the lower LFT are ordered by $1$, $u$, and $w$.  This stacks $u=w_0$ directly above $w=[(w^1)^\top,\ldots, (w^L)^\top]^\top$, i.e. $w^0$ is ordered above $w^1$.  This ordering is the reason to choose the lower LFT over the more common use of an upper LFT for uncertain systems. This example illustrates the flexibility of the LFT notation for modeling NNs. The results that follow in the paper hold for more general implicit NNs modeled in the LFT formulation~\eqref{eq:LFT}.

\section{Background}\label{sec:back}

This section reviews an existing method, LipSDP, to compute a bound on the Lipschitz constant for a NN \cite{Fazlyab2019,pauli21training}. The approach uses quadratic constraints to characterize the incremental input/output behavior of the nonlinearity $\Phi$ \cite{megretski97system,Fazlyab2022}. This leads to a linear matrix inequality (LMI) condition that is sufficient to bound the Lipschitz constant.  

\subsection{Quadratic Constraints (QCs)}

The nonlinearity $\Phi$ is the main complication in assessing the Lipschitz constant of a neural network. It is useful to characterize the input/output properties of this nonlinearity by a simple, but possibly conservative, quadratic constraint. This is formalized in the next definition.
\vspacelem
\begin{defn}
A function $\Phi:\R^m\rightarrow \R^m$ satisfies the quadratic constraint defined by $M\in\Sy^{2m}$ if
\begin{equation}
\bmtx v \\ w \emtx^\top M \bmtx v \\ w \emtx \ge 0
\hspace{0.2in}
\forall v\in \R^m, \,\, w=\Phi(v).
\end{equation}
\end{defn}
\vspacelem

The matrix $M$ is called the multiplier of the quadratic constraint. We define~$QC(M)$ as the set of all functions that satisfy the QC defined by~$M$, i.e., $\Phi\in QC(M)$ if and only if $\Phi$ satisfies the QC defined by~$M$.

Common NN activation functions have properties that can be expressed as QCs. For example, most activation functions lie in the sector $[\alpha,\beta]$ for some $\beta\ge \alpha$, i.e.,
\begin{align}
\label{eq:sc1}
(\beta v - w) \, (w-\alpha v) \ge 0
\hspace{0.2in} \forall v\in \R^m, \, w=\phi(v).
\end{align}
In words, the graph of the activation function satisfies $\phi(0)=0$ and lies between lines of slope $\alpha$ and $\beta$. In addition, most NN activation functions are slope restricted. Specifically a function $\phi$ is said to be slope restricted to $[\alpha,\beta]$ for some $\beta\ge \alpha$ if $\phi(0)=0$ and:
\begin{equation}\label{eq:sr1}
    \alpha \le  \frac{\phi(\bar{v}) - \phi(\hat{v})}{\bar{v}-\hat{v}}\le \beta\quad 
    \forall\,\bar{v},\,\hat{v}\in \R^m
    \mbox{ with } \bar{v} \ne \hat{v}
\end{equation}
The slope bounds can be equivalently written as a single quadratic inequality:
\begin{align}\label{eq:sr2}
\begin{split}    
    \left( \beta(\bar{v}-\hat{v})-(\bar{w}-\hat{w}) \right) \,
    \left( (\bar{w}-\hat{w})-\alpha(\bar{v}-\hat{v}) \right) 
    \ge 0 \\
    \forall\,\bar{v},\,\hat{v}\in \R^m, \,
    \bar{w}=\phi(\bar{v}),\,
    \hat{w}=\phi(\hat{v})
\end{split}
\end{align}
Note that if a function is slope restricted to $[\alpha,\beta]$ then it also satisfies the sector bounds $[\alpha,\beta]$.
The next two lemmas define sets of QCs for a repeated nonlinearity $\Phi$ that is sector bounded or slope restricted.

\vspacelem
\begin{lem}\label{lem:qc_sb}
Consider a repeated nonlinearity $\Phi:\R^m\to\R^m$ constructed from an activation function $\phi:\R\to\R$ in the sector $[\alpha,\beta]$. Define the following subset of $\Sy^{2m}$
\begin{equation}
\Msec(\alpha,\beta) := \left\{\bmtx -(2\alpha\beta) \, \Lambda & (\alpha +\beta) \, \Lambda  \\  (\alpha +\beta) \, \Lambda & -2\Lambda \emtx: \Lambda \in \D_{\ge 0}^m\right\}.
\end{equation}
Then $\Phi$ satisfies the QC defined by any $M\in\Msec(\alpha,\beta)$.
\end{lem}
\begin{proof}
This is a well-known condition for sector-bounded functions (Section 8.1 of\cite{boyd94linear}
 or Section VI.I of \cite{megretski97system}).
If $M\in\Msec(\alpha,\beta)$ then for any $v$ and $w=\Phi(v)$,
\begin{align*}
    \bmtx v \\ w \emtx^\top M \bmtx v \\ w \emtx  = 
\sum_{i=1}^m 2 \Lambda_{ii}
(\beta v_i - w_i) \, (w_i-\alpha v_i) 
\end{align*}
The sum is nonnegative by the sector
condition~\eqref{eq:sc1} and
$\Lambda \in \D_{\ge 0}^m$, 
\end{proof}
\vspacelem

The next lemma provides a richer class of multipliers formulated with respect to a doubly hyperdominant matrix, see, e.g., \cite{Ebihara2024local, fetzer2017absolute}.
\begin{lem}\label{lem:qc_sr}
Consider a repeated nonlinearity $\Phi:\R^m\to\R^m$ constructed from an activation function $\phi:\R\to\R$ slope restricted to $[\alpha,\beta]$. Define the following subset of $\Sy^{2m}$
\begin{align}
& \Mslope(\alpha,\beta) 
:= \\
\nonumber
& \left\{  
\bmtx -(\alpha\beta)\, (Q+Q^\top) & \alpha \, Q +\beta \, Q^\top  \\  \alpha \, Q^\top  +\beta \, Q & -(Q+Q^\top) \emtx 
\, : \, Q \in \DHyp^m\right\}.
\end{align}
Then $\Phi$ satisfies the QC defined by any $M\in\Mslope(\alpha,\beta)$.
\end{lem}
\begin{proof}
If $M\in\Mslope(\alpha,\beta)$ then for any $v$ and $w=\Phi(v)$,
\begin{align}
\label{eq:scproof}
    \bmtx v \\ w \emtx^\top M \bmtx v \\ w \emtx   
    = 2\, (w-\alpha v)^\top Q (\beta v-w).
\end{align}
It follows from the results in Section 3.5 of \cite{willems71} that this expression is nonnegative. Specifically, define $\tilde{v} :=\beta v-w$ and $\tilde{w} := w-\alpha v$. The slope condition~\eqref{eq:sr2} implies that the data $\{ (\tilde{v}_i, \tilde{w}_i) \}_{i=1}^m$ is similarly ordered, i.e. $\tilde{v}_k > \tilde{v}_j$ implies  $\tilde{w}_k \ge \tilde{w}_j$. Moreover, slope-restricted functions are also sector bounded.  It follows from the sector condition~\eqref{eq:sc1} that the data is unbiased, i.e.  $\tilde{v}_k \tilde{w}_k  \ge 0$ for $k=1,\ldots,m$. Theorem 3.8 in~\cite{willems71} implies that if $Q$ is doubly hyperdominant then $\tilde{w}^\top Q \tilde{v} \ge 0$, i.e. the expression in \eqref{eq:scproof}
is nonnegative.
\end{proof}
\vspacelem

Note that $\Lambda$ is restricted to be a diagonal, nonnegative matrix in the set $\Msec(\alpha,\beta)$. However, $Q$ can be any (possibly full / non-symmetric) doubly-hyperdominant matrix in the set $\Mslope(\alpha,\beta)$. The slope-restriction property enlarges the set of valid QCs for a function. Other QCs can be constructed that exploit additional properties of the activation function. For example, classes of QCs have been developed that exploit specific properties of repeated ReLU \cite{richardson23,drummond24,fazlyab2020safety,ebihara21EJC,ebihara21CDC,noori2024stability,noori2026dynamic}.  The complete set of of all possible QCs for repeated ReLU was derived in~\cite{Noori2025}. This complete set, denoted $\Mcrelu$, is characterized by a collection of copositivity conditions on the matrix~$M$. The complete set defines the largest possible class of QCs for repeated ReLU.

\subsection{Incremental QCs}

We can also define QCs on the increments of the input and output $\bar{v} - \hat{v}$ and $\Phi(\bar{v}) - \Phi( \hat{v} )$, respectively. A formal definition for such incremental QCs is provided next.

\vspacelem
\begin{defn}
A function $\Phi:\R^m\rightarrow \R^m$ satisfies the incremental quadratic constraint defined by $M\in\Sy^{2m}$ if
\begin{align}\label{eq:qcinc}
\begin{split}
& \bmtx \bar{v} -\hat{v} \\ \bar{w}-\hat{w}\emtx^\top M \bmtx \bar{v}-\hat{v} \\ \bar{w}-\hat{w} \emtx \ge 0 \\
& \forall \bar{v}, \, \hat{v}\in \R^m, \,\, \bar{w}=\Phi(\bar{v}), \,\,
\hat{w}=\Phi(\hat{v}).
\end{split}
\end{align}
\end{defn}
\vspacelem

Again, $M$ is called the multiplier of the incremental QC. We define~$QC^\text{inc}(M)$ as the set of all functions that satisfy the incremental QC defined by~$M$. The next lemma defines a simple class of incremental QCs for a repeated, slope-restricted nonlinearity. 
\vspacelem
\begin{lem}\label{lem:incqc_sr}
Consider a repeated nonlinearity $\Phi:\R^m\to\R^m$ constructed from an activation function $\phi:\R\to\R$ slope restricted to $[\alpha,\beta]$. Define the following subset of $\Sy^{2m}$
\begin{equation}
\Mislope(\alpha,\beta) 
:= \left\{\bmtx -(2\alpha\beta)\, \Lambda & (\alpha +\beta) \, \Lambda  \\  (\alpha +\beta) \,\Lambda & -2\Lambda \emtx: \Lambda \in \D_{\ge 0}^m\right\}
\end{equation}
Then $\Phi$ satisfies the incremental QC defined by any $M\in\Mislope(\alpha,\beta)$.
\end{lem}
\begin{proof}
Again, this is a well-known condition for slope restricted functions \cite{willems68some,lessard16analysis}.
Let any $\bar{v}$ and $\hat{v}\in\R^m$ be given with corresponding outputs  $\bar{w}=\Phi(\bar{v})$, $\hat{w} = \Phi(\hat{v})$. Define the increments $\Delta v=\bar{v}-\hat{v}$
and $\Delta w = \bar{w} - \hat{w}$. If $M\in\Mislope(\alpha,\beta)$ then 
\begin{align*}
& \bmtx \Delta v \\ \Delta w \emtx^\top M 
  \bmtx \Delta v \\ \Delta w \emtx  =  \\
& \hspace{0.1in} \sum_{i=1}^m 2 \Lambda_{ii}
\left(\beta (\Delta v)_i - (\Delta w)_i \right) \, 
\left( (\Delta w)_i-\alpha (\Delta v)_i \right) 
\end{align*}
The sum is nonnegative by the slope condition~\eqref{eq:sr2} and $\Lambda \in \D_{\ge 0}^m$, \end{proof}
\vspacelem

Note that the set $\Mislope(\alpha,\beta)$ is the same 
as the set $\Msec(\alpha,\beta)$. The former defines incremental QCs for slope-restricted functions while the latter defines (non-incremental) QCs for sector-bound functions. As for the non-incremental case, there are incremental QCs that exploit specific properties of specific activation functions to reduce conservatism. For example, the complete set of incremental QCs for ReLU, denoted $\Micrelu$, is presented in~\cite{Noori2025}.

\subsection{Global Lipschitz Bounds}

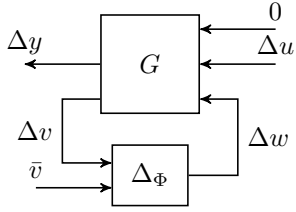
\begin{figure}[ht!]
\centering
\begin{tikzpicture}[blockdiag]
	\node[block,minimum width=1.3cm, minimum height=1.3cm](Plant) {$G$};
	\node[block, below= 0.4cm of Plant] (Delta) {$\Delta_\Phi$};

	\draw[<-] ($(Plant.south east)!.5!(Plant.north east)$) -- +(+1.0cm, 0) 
	node[above, name=e]{$\Delta u$};

\draw[<-] ($(Plant.south east)!.85!(Plant.north east)$) -- +(+1.0cm, 0) 	node[above, name=e]{$0$};

    \draw[->] ($(Plant.south west)!.5!(Plant.north west)$) -- +(-1.0cm, 0) 
	node[above, name=d]{$\Delta y$};
	
	\draw[<-] ($(Plant.south east)!.15!(Plant.north east)$) -- +(+0.5cm, 0) |- 
	node[near start, right]{$\Delta w$} (Delta);	
	\draw[->] ($(Plant.south west)!.15!(Plant.north west)$) -- +(-0.5cm, 0) |- 
	node[near start, left ]{$\Delta v$} ($(Delta.west)!0.4!(Delta.north west)$);

	\draw[<-] ($(Delta.west)!0.4!(Delta.south west)$) -- +(-1.0cm, 0)
	node[above]{$\bar{v}$};
\end{tikzpicture}
\caption{Incremental system for Lipschitz bound analysis.}
\label{fig:UncSysDelta}
\vspace{0pt}
\end{figure}

The LipSDP framework~\cite{Fazlyab2019,pauli21training} provides a sufficient condition to calculate an upper bound~$\gamma$ on the Lipschitz constant of a given NN. This framework is based on an incremental representation of the NN as shown in Figure~\ref{fig:UncSysDelta}. Specifically, assume the NN is well-posed and consider any input $\bar{u}$. Let $(\bar{v},\bar{w},\bar{y})$ be the resulting vectors that satisfy the LFT relations in Equation~\eqref{eq:LFT}. Similarly, let $\hat{u}$ be another input with solutions $(\hat{v},\hat{w},\hat{y})$. The increments are defined as the difference between these two solutions, e.g. $\Delta u := \bar{u} - \hat{u}$. These increments satisfy the following equations:
\begin{subequations}
\label{eq:incLFT}
\begin{align}
\label{eq:incLFTy}
        \Delta y &=  G_{12}\, \Delta u + G_{13}\, \Delta w \\
\label{eq:incLFTv}
        \Delta v &= G_{22}\, \Delta u, + G_{23}\, \Delta w  \\
\label{eq:incLFTw}
        \Delta w & = \Phi(\bar{v}) -\Phi(\hat{v}).
\end{align}
\end{subequations}
Equations~\eqref{eq:incLFTy} and~\eqref{eq:incLFTv} are expressed entirely in terms of the increments. This follows because the known part $G$ defines an affine relation between the vectors. Note that the bias terms $G_{11}$ and $G_{21}$ cancel out in this incremental form.  

The main difficulty lies in Equation~\eqref{eq:incLFTw}. The activation function is nonlinear and hence $\Delta w$ is not simply a function of $\Delta v$, in general. Instead, $\Delta w$ depends independently on $\bar{v}$ and $\hat{v}$. Alternatively, we can equivalently write $\Delta w$ as a function of $\bar{v}$ and $\Delta v:=\bar{v}-\hat{v}$. In particular, define $\Delta_\Phi:\R^m \times \R^m \to \R^m$ as follows:
\begin{align}
\label{eq:incLFTw2}
    \Delta w= \Delta_\Phi (\Delta v,\bar{v})
            := \Phi(\bar{v}) -\Phi(\bar{v}-\Delta v).
\end{align}
In summary, we can represent the incremental input/output relation for the NN by the LFT equations \eqref{eq:incLFTy},
\eqref{eq:incLFTv}, and \eqref{eq:incLFTw2}. This corresponds to the interconnection shown in~Figure~\ref{fig:UncSysDelta}. The zero entering in the first input of $G$ indicates that the bias terms are removed from the nominal part. It is important to emphasize that the bias terms still affect the mapping from $\Delta u$ to $\Delta y$ due to their impact on $\bar{v}$.

The LipSDP framework bounds the Lipschitz constant of the NN based on this incremental system. The condition is formally stated in Theorem~\ref{thm:lipsdp} below. The condition uses the affine function $L_\text{Lip}: \Sy^{2m} \times \R_{\ge0} \rightarrow \Sy^{m+n_u}$ defined as:
\begin{align}
\nonumber
& L_\text{Lip}(M, \gamma^2) 
 := \bmtx G_{22} & G_{23} \\ 0  & I_m \emtx^\top M 
  \bmtx G_{22} & G_{23} \\ 0  & I_m \emtx\\
\label{eq:lipLMI}
& \hspace{0.3in}
+ \bmtx G_{12}^\top \\ G_{13}^\top \emtx \, 
        \bmtx G_{12} & G_{13} \emtx
 -\bmtx \gamma^2 I_{n_u}  & 0 \\ 0 & 0_{m}\emtx 
\end{align}
The LipSDP condition with the incremental system is formally stated in Theorem~\ref{thm:lipsdp} below.
\vspacelem
\begin{mytheo}\label{thm:lipsdp}
Consider a well-posed NN $f:\R^{n_u} \to \R^{n_y}$ defined by the LFT $F_\text{l}(G,\Phi)$ in  Equation~\ref{eq:LFT}. Assume that the repeated nonlinearity $\Phi:\R^m\to \R^m$ satisfies the incremental QC defined by $M\in \Sy^{2m}$. Further assume that there exists a scalar $\gamma \ge 0$ such that $L_\text{Lip}(M,\gamma^2)\preceq 0$. Then the Lipschitz constant of the NN $f$ is bounded by $\gamma$.
\end{mytheo}
\begin{proof}
This follows directly from Theorem 2 in \cite{Fazlyab2019}. We briefly sketch the proof for completeness. Let $\bar{u}$, $\hat{u} \in \R^{m}$ be two inputs to the NN. The NN is assumed to be well-posed, so there exist corresponding
solutions $(\bar{v},\bar{w},\bar{y})$ and 
$(\hat{v},\hat{w},\hat{y})$. Multiply $L_\text{Lip}(M,\gamma^2)\preceq 0$ on the left by $[\Delta u^\top, \, \Delta w^\top]$ and on the right by its transpose. The incremental LFT equations \eqref{eq:incLFTy} and \eqref{eq:incLFTv} then give:
\begin{align}
\label{eq:LipSDIneq}
\bmtx \Delta v \\ \Delta w \emtx^\top M 
  \bmtx \Delta v \\ \Delta w \emtx  
  + \Delta y ^\top \Delta y - \gamma^2 \Delta u^\top \Delta u \le 0
\end{align}
The first term is non-negative because $\Phi$ satisfies the incremental QC defined by $M$. 
This yields $\| \Delta y\|_2 \le \gamma \| \Delta u \|_2$, hence the NN Lipschitz constant is upper bounded by $\gamma$.
\end{proof}
\vspacelem
This theorem provides a sufficient condition to bound the Lipschitz constant of the NN.  Suppose the activation satisfies any QC defined by a multiplier $M$ in the set $\mcl{M}\subset \Sy^{2m}$. We can solve the following optimization to find the best (smallest) bound on the Lipschitz constant:
\begin{align}
\label{eq:IncSDP}
    \min_{M\in \mcl{M}} \gamma^2
    \mbox{ s.t. } L_\text{Lip}(M,\gamma^2) \preceq 0.
\end{align}
The constraint $L_\text{Lip}(M,\gamma^2) \preceq 0$ is a linear matrix inequality (LMI) in $M$ and $\gamma^2$. Moreover, the set $\mcl{M}$ if often described by affine constraints on $M$. In this case, the optimization \eqref{eq:IncSDP} is a (convex) semidefinite program (SDP).

Also note that we assumed the NN is well-posed in Theorem~\ref{thm:lipsdp}.  However, well-posedness follows directly from a strict version of the LMI condition in certain cases (and does not need to be assumed). Consider activation functions that are slope restricted to $[0,1]$. By Lemma~\ref{lem:incqc_sr}, the function $\Phi$ satisfies the incremental QC defined by any multiplier in the set:
\begin{align}
\label{eq:Msrfor01}
\Mislope(0,1)=\left\{ M = \bmtx 0 & \Lambda \\ \Lambda & -2\Lambda \emtx \, : \,
\Lambda \in \D_{\ge 0}^m
\right\}
\end{align}
Suppose $L_\text{Lip}(M,\gamma^2)\prec 0$ is feasible for some multiplier $M$ in this set. The lower right block of this LMI then implies:
\begin{align}
    2\Lambda - \Lambda G_{23} - G_{23}^\top \Lambda
    \succ G_{13}^\top G_{13} \succeq 0.
\end{align}
It follows from Theorem 1 in \cite{revay20lipschitz} that the NN is well-posed.

\section{Main Results}\label{sec:main}

The existing LipSDP condition in Theorem~\ref{thm:lipsdp} uses the incremental LFT representation (Figure~\ref{fig:UncSysDelta}) with an incremental QC specified on $\Delta v$ and $\Delta w$. The LMI condition in this framework does not involve the bias terms.  Moreover, incremental QCs on $(\Delta v, \Delta w)$ do not capture all the possible QCs that couple $(\bar{v},\hat{v},\bar{w},\hat{w})$. In Section~\ref{sec:motivation} we show that these two issues introduce conservatism in the analysis. This motivates our alternative formulation introduced in the remainder of the section.

\subsection{Motivation}
\label{sec:motivation}

\begin{figure}[ht!]
\centering
\begin{tikzpicture}[>=stealth]
\definecolor{blue1}{RGB}{222,235,247}
\definecolor{blue2}{RGB}{158,202,225}
\definecolor{blue3}{RGB}{49,130,189}

\begin{axis}[ width = 0.65\columnwidth, height = 0.65\columnwidth, 
    name=leftplot,
    grid=major,
    xlabel=  $u$, 		
    ylabel=  $y$, 	
    xmin = -2.25, xmax = 2.25, 
    ymin = -10.25, ymax = 6, 
    axis lines=middle,   
    xtick={-2.0,-1.0,1,0, 2},
    ytick={-10,-8, -6,-4.0, -2.0,2, 4},
]


\addplot[blue3, dotted, line width=2.5pt] coordinates {(-2.5,0) (0,0) (2,-4)};\label{pl:ex2}
\addplot[blue3, line width=2.5pt] coordinates {(-2.5,0) (-1,0) (1,4) (2,2)};\label{pl:ex1}

\addplot[blue3, loosely dashdotted, line width=2.5pt] coordinates {(-2.5,0) (-1,0) (1,-8) (2,-10)};\label{pl:ex3}

\end{axis}

\end{tikzpicture}
\caption{Input-output behavior with three different biases:  (a) $b^1 = [1 \, -1]^\top$ (\ref{pl:ex1}), (b) $b^1 = [0 \, 0]^\top$ (\ref{pl:ex2}), and (c) $b^1 = [-1 \, 1]^\top$ (\ref{pl:ex3}).}
\label{fig:ex1}
\vspace{0pt}
\end{figure}
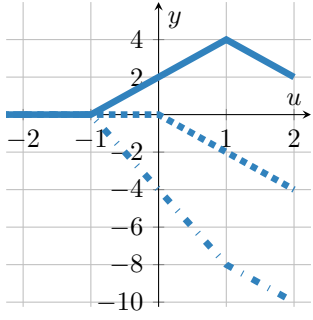

We first present a simple example to illustrate the effect of neglecting the bias terms. Our example is inspired by, but different from, the
example in Section VIII.A of \cite{gronqvist22}.
Consider an MLP in \eqref{eq:NNcomp} mapping $u\in\R$ to $y\in\R$ with a single layer $(L=1)$ and two neurons ($m=2$).  This is represented as a lower LFT with the matrix $G$ defined by:
\begin{align}
& G_{11} = b^2, \,\, G_{12} = 0, \,\, G_{13} = W^2, \\
& G_{21} = b^1, \,\, G_{22} = W^1, \,\, G_{23} = 0.
\end{align}
The NN weights and biases for the example are:
\begin{align*}
    & W^1 = \bmtx 1 \\ 1\emtx,  \,
    W^2 =\bmtx 2 & -4\emtx, \,
    b^1=\bmtx 1 \\  -1\emtx, \mbox{ and } 
    b^2=0.
\end{align*}
The nonlinearity $\Phi:\R^2\to \R^2$ is a repeated ReLU, i.e. the scalar activation function is $\phi(v)=\max(v,0)$. The corresponding input-output map for this NN is:
\begin{equation}\label{eq:ffnobias2}
    y = \bmtx 2 & -4\emtx \Phi\left(\bmtx 1 \\ 1 \emtx u +\bmtx 1 \\ -1 \emtx \right),
\end{equation}
Figure~\ref{fig:ex1} shows the graph of this function, labeled as case (a). The function
transitions between slopes of 0, +2, and finally -2. Hence the true Lipschitz constant is $\gamma_\text{true} = 2$.

We can bound the Lipschitz constant for this NN using the LipSDP condition in  Theorem~\ref{thm:lipsdp}. ReLU is a $[0,1]$ slope-restricted function. Hence, it satisfies all incremental QCs defined by multipliers $\Mislope(0,1)$ as in \eqref{eq:Msrfor01}. We can solve the following SDP to find the best (smallest) bound on the Lipschitz constant:
\begin{align}\label{eq:lip}
    \min_{M\in \Mislope(0,1)} \gamma^2
    \mbox{ s.t. } L_\text{Lip}(M,\gamma^2) \preceq 0.
\end{align}
Solving this SDP gives the Lipschitz bound $\gamma=4$. For this example, the optimum can also be obtained analytically using the slope-restricted multiplier, with $\lambda_1=8$ and $\lambda_2=16$. We also solved the SDP  using the complete set for the repeated ReLU, $\Micrelu$ \cite{Noori2025}. The complete set contains every valid multiplier for repeated ReLU. Hence, this set gives the least conservative bound achievable within the LipSDP framework. However, even in this case the LipSDP bound is $\gamma=4$. This bound is conservative: $\gamma>\gamma_\text{true}=2$.

The first source of conservatism arises because the LipSDP condition in Theorem~\ref{thm:lipsdp} does not involve the bias terms.   Figure~\ref{fig:ex1} shows the graph of this function for two additional biases: 
(b) $b^1 = [0 \, 0]^\top$ and (c) $b^1=[-1 \, 1]^\top$. In the bias-free case (b), the function transitions from a slope of 0 to a slope of $-2$. The Lipschitz constant is thus $\gamma_\text{true}=2$ for this case. However, the function with bias (c) transitions between slopes of $0$, $-4$, and finally $-2$. Thus the Lipschitz constant for bias (c) is $\gamma_\text{true}=4$. 
The LipSDP condition does not involve the bias and hence yields the same bound for all choices of the bias. Thus, the Liptchitz bound can be conservative for any specific NN biases.

We next consider a second source of conservatism, namely, the restriction for a repeated function $\Phi:\R^m\to \R^m$ to QCs that are solely expressed in terms of the increments $(\Delta v, \Delta w)$. Specifically, the incremental QC \eqref{eq:qcinc} defines a constraint on the increments $\Delta v = \bar{v} - \hat{v}$
and $\Delta w = \bar{w} - \hat{w}$. This can be 
equivalently written as a QC on the original variables $(\bar{v},\hat{v},\bar{w},\hat{w})$. To see this, define the stacked inputs and outputs:
\begin{align}
\label{eq:vewe}
v^\text{s} := \bmtx \bar{v} \\ \hat{v} \emtx \in\R^{2m}
\mbox{ and }
w^\text{s} := \bmtx \bar{w} \\ \hat{w} \emtx \in\R^{2m}.
\end{align}
The incremental QC with multiplier $M$ can then be rewritten in terms of these stacked vectors:
\begin{align}
\label{eq:incQCext}
0 \le 
\bmtx \Delta v \\ \Delta w \emtx^\top 
M
\bmtx \Delta v \\ \Delta w \emtx
=
\bmtx v^\text{s} \\ w^\text{s} \emtx^\top
T_\Delta^\top M T_\Delta
\bmtx v^\text{s} \\ w^\text{s} \emtx,
\end{align}
where $T_\Delta \in \R^{2m\times 4m}$ is:
\begin{align}
T_\Delta := \bmtx I_m & -I_m & 0 & 0 \\ 
0 & 0 & I_m & -I_m \emtx.
\end{align}
Let $\Phi^\text{s}:\R^{2m}\to\R^{2m}$ denote the 
the stacked two-copy nonlinearity
that maps $v^\text{s}$ to $w^\text{s}$. The right side of \eqref{eq:incQCext} is in the form of a standard (non-incremental) QC on the stacked function $\Phi^\text{s}$. Thus if $M\in \Sy^{2m}$ defines an incremental QC for $\Phi$ then $T_\Delta^\top M T_\Delta\in \Sy^{4m}$ defines a (non-incremental) QC for $\Phi^\text{s}$.

The important point is that there are (non-incremental) QCs that can be defined for $\Phi^\text{s}$ that do not correspond to an incremental QC for $\Phi$. To illustrate this, consider the special case where $\Phi:\R^m \to \R^m$  is $[0,1]$ slope-restricted. 
By Lemma~\ref{lem:incqc_sr}, $\Phi$
satisfies any incremental QC defined by a multiplier 
$M\in \Mislope(0,1)$. These multipliers have the form:
\begin{align}
M = \bmtx 0 & \Lambda \\ \Lambda & -2\Lambda \emtx \, \mbox{ where }
\Lambda \in \D_{\ge 0}^m.
\end{align}
The corresponding (non-incremental) QCs on the stacked function $\Phi^\text{s}$ have the form:
\begin{align}
\label{eq:RMR01}
T_\Delta ^\top M T_\Delta = \bmtx 0 & Q \\ Q & -2Q \emtx \, \mbox{ where }
Q:= \bmtx \Lambda & -\Lambda \\ -\Lambda & \Lambda \emtx.
\end{align}
However, there is a larger class of valid QCs for the stacked function.  By Lemma~\ref{lem:qc_sr}, $\Phi^\text{s}:\R^{2m}\to\R^{2m}$ satisfies any QC defined by a multiplier in the following set:
\begin{align}
\Mslope(0,1) 
:= 
\left\{  
\bmtx 0 & Q^\top  \\   Q & -(Q+Q^\top) \emtx 
\, : \, Q \in \DHyp^{2m}\right\}.
\end{align}
There are many possible choices for $Q$ in the set $\Mslope(0,1)$ that do not have the special block structure in \eqref{eq:RMR01}.  In other words, incremental QCs on $(\Delta v, \Delta w)$ do not capture all the possible QCs that couple $(\hat{v},\bar{v},\hat{w},\bar{w})$.
This is important as larger sets of QCs can
reduce conservatism in the analysis.

\subsection{Lipschitz Bounds With Extended LFT System}

\begin{figure}[ht!]
\centering
\begin{tikzpicture}[blockdiag]
	
\node[block,minimum width=1.3cm, minimum height=1.3cm](Plant) {$G$};
\node[block, below= 0.4cm of Plant] (Delta) {$\Phi$};
\draw[<-] ($(Plant.south east)!.5!(Plant.north east)$) -- +(2.2cm,0) node[above, name=e]{$\bar{u}$};
\draw[<-] ($(Plant.south east)!.2!(Plant.north east)$) -- +(+0.5cm, 0) |-  node[near start, right]{$\bar{w}$} (Delta);	
\draw[->] ($(Plant.south west)!.2!(Plant.north west)$) -- +(-0.5cm, 0) |- node[near start, left ]{$\bar{v}$} (Delta);

\node[block,minimum width=1.3cm, minimum height=1.3cm, below= 1.5cm of Plant] (Plant2) {$G$};
\node[block, below= 0.4cm of Plant2] (Delta2) {$\Phi$};
\draw[<-] ($(Plant2.south east)!.5!(Plant2.north east)$) -- +(2.2cm,0) node[above, name=e]{$\hat{u}$};
\draw[<-] ($(Plant2.south east)!.2!(Plant2.north east)$) -- +(+0.5cm, 0) |- node[near start, right]{$\hat{w}$} (Delta2);	
\draw[->] ($(Plant2.south west)!.2!(Plant2.north west)$) -- +(-0.5cm, 0) |- node[near start, left ]{$\hat{v}$} (Delta2);

\node[sum, left= 1.4cm of Plant2, yshift = 0.26cm](Sum1){};
\draw[->] ($(Plant.south west)!.7!(Plant.north west)$)node[above, name=d, xshift = -0.5cm]{$\bar{y}$} -| (Sum1.north);
\draw[->] ($(Plant2.south west)!.7!(Plant2.north west)$) node[above, name=d, xshift = -0.5cm]{$\hat{y}$} -- +(Sum1);
\draw[->] (Sum1.west) -- +(-1.0cm, 0) node[above, name=d]{$\Delta y$};
\node[right=0.1cm of Sum1, yshift = 0.15cm]{$-$};

\node[connector, right= 1.4cm of Plant, yshift = 0.40cm](Con1){};
\draw[<-] ($(Plant.south east)!0.8!(Plant.north east)$) -- (Con1);
\draw[<-] ($(Plant2.north east)!.3!(Plant2.east)$) -| +(Con1);
\draw[-] (Con1) -- +(+0.8cm, 0)  node[above, name=e]{$1$};

\end{tikzpicture}
\caption{Stacked system for Lipschitz bound analysis.}
\label{fig:UncSysNew}
\vspace{0pt}
\end{figure}
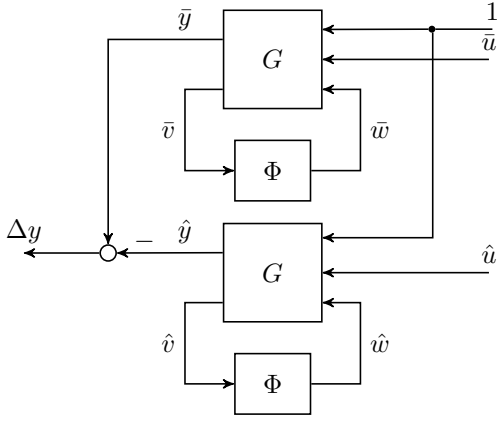

We use the two-copy LFT representation in Figure~\ref{fig:UncSysNew} for our incremental analysis. This system consists of two copies of the original LFT $F_\text{l}(G,\Phi)$ driven by the independent exogenous inputs $\bar{u}$ and $\hat{u}$, while sharing the constant input $1$ associated with the bias terms. The output of the stacked system is $\Delta y = \bar{y} - \hat{y}$. Our goal is to show that this stacked system satisfies $\|\Delta y\|_2 \le \gamma \|\Delta u\|_2$ for all input pairs $(\bar{u},\hat{u})$ with $\Delta u=\bar{u}-\hat{u}$. This corresponds to a Lipschitz bound $\gamma$ on the original LFT $F_\text{l}(G,\Phi)$.

\begin{figure}[ht!]
\centering
\begin{tikzpicture}[blockdiag]
	
\node[block,minimum width=1.3cm, minimum height=1.3cm](Plant) {$G^\text{e}$};
\node[block, below= 0.4cm of Plant] (Delta) {$\Phi^\text{e}$};

\draw[<-] ($(Plant.south east)!.5!(Plant.north east)$) -- +(+1.0cm, 0) 
	node[above, name=e]{$u^\text{e}$};

\draw[<-] ($(Plant.south east)!.85!(Plant.north east)$) -- +(+1.0cm, 0) 
	node[above, name=e]{$1$};
	
\draw[->] ($(Plant.south west)!.5!(Plant.north west)$) -- +(-1.0cm, 0) 
	node[above, name=d]{$\Delta y$};
	
\draw[<-] ($(Plant.south east)!.15!(Plant.north east)$) -- +(+0.5cm, 0) |- node[near start, right]{$w^\text{e}$} (Delta);	
\draw[->] ($(Plant.south west)!.15!(Plant.north west)$) -- +(-0.5cm, 0) |- node[near start, left ]{$v^\text{e}$} (Delta);

\end{tikzpicture}

\caption{Representation of stacked system as an interconnection $F_\text{l}(G^\text{s},\Phi^\text{s})$ of a matrix $G^\text{s}$ and nonlinearity $\Phi^\text{s}$.}
\label{fig:UncSys3}
\vspace{0pt}
\end{figure}
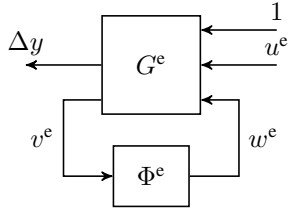

This stacked system can be expressed as an interconnection of a stacked matrix $G^\text{s}$ and a stacked nonlinearity $\Phi^\text{s}$ as shown in Figure~\ref{fig:UncSys3}. The vectors $v^\text{s}$ and $w^\text{s}$ were defined in \eqref{eq:vewe} by stacking $(\bar{v},\hat{v})$ and $(\bar{w},\hat{w})$, respectively.  Similarly, the stacked input $u^\text{s}\in \R^{2m}$ is defined by stacking $(\bar{u},\hat{u})$.
The stacked nonlinearity $\Phi^\text{s}:\R^{2m}\to\R^{2m}$ maps  $w^\text{s} = \Phi^\text{s}(v^\text{s})$. This is a repeated nonlinearity with $2m$ copies of the scalar activation function $\phi$.

Figure~\ref{fig:UncSys3} corresponds to the following  equations:
\begin{subequations}
\label{eq:eLFT}
\begin{align}
\label{eq:eLFTy}
 \Delta  y &= 
G_{11}^\text{s}
+ G_{12}^\text{s}\,  u^\text{s}
+ G_{13}^\text{s}\,  w^\text{s}, \\
\label{eq:eLFTv}
v^\text{s} &= 
G_{21}^\text{s}
+ G_{22}^\text{s}\,  u^\text{s}
+ G_{23}^\text{s}\,  w^\text{s}, \\
\label{eq:eLFTw}
w^\text{s}
& = \Phi^\text{s}\left( v^\text{s} \right).
\end{align}
\end{subequations}
  The stacked matrix $G^\text{s}$ 
can be constructed from the original matrix $G$ with the following
block definitions:
\begin{align*}
    G_{11}^\text{s} & := 0, \,\,
    G_{12}^\text{s} := \bmtx G_{12} & -G_{12}\emtx,  G_{13}^\text{s} := \bmtx G_{13} & -G_{13} \emtx,
\\
    G_{21}^\text{s} & := \bmtx G_{21} \\ G_{21} \emtx, 
    G_{22}^\text{s} :=  \bmtx G_{22}  & 0 \\ 0 & G_{22}\emtx,
    G_{23}^\text{s} := \bmtx G_{23} & 0 \\ 0 & G_{23}\emtx.
\end{align*}

The two-copy LFT representation~$F_\text{l}(G^\text{s},\Phi^\text{s})$ of the stacked system contains all
signals required for the Lipschitz analysis of the NN.
In particular, this stacked system contains the bias term $G_{21}$.\footnote{The output bias term $G_{11}$ does not appear as it cancels out in the expression for $\Delta y$. Hence  $G_{11}$ has no effect on the Lipschitz constant.} This is in contrast to the incremental system (Figure~\ref{fig:UncSysDelta}) used in standard Lipschitz bound analyses, which does not contain the bias terms.  Moreover, the stacked system contains the vectors $(v^\text{s},w^\text{s})$ needed to write all possible QCs for $\Phi^\text{s}$. This enables a larger class of constraints than can be expressed with incremental QCs in the standard analysis.

\subsection{Bias Quadratic Constraints}

The QCs introduced in Section~\ref{sec:back} are defined for~$[v^\top,w^\top]^\top$ or in the incremental case~$[\Delta v^\top, \Delta w^\top]^\top$. They lack affine and constant terms.  Thus, they cannot explicitly capture affine, bias-dependent information associated with the term~$G_{21}^\text{s}$. Biased QCs \cite{materassi09, fazlyab2019efficient} present a richer class of quadratic constraints, which are defined in terms of the vector~$[1, v^\top,w^\top]^\top$ and a corresponding matrix~$M^\text{b}$. This augmentation allows the quadratic form to include both affine and constant terms. We present a formal definition of such a biased QC next.

\vspacelem
\begin{defn}
\label{def:biasedQC}
A function $\Phi:\R^m\rightarrow \R^m$ satisfies the biased quadratic constraint defined by $M^\text{b}\in\Sy^{2m+1}$ if
\begin{equation}
\label{eq:biasedQC}
\bmtx 1\\ v \\ w \emtx^\top M^\text{b} \bmtx 1 \\ v \\ w\emtx \ge 0
\hspace{0.2in}
\forall v\in \R^m, \,\, w=\Phi(v).
\end{equation}
\end{defn}
\vspacelem

Several different biased QCs for general nonlinearities were proposed in literature covering, for example, nonlinearities confined
to hyperbolic sectors~\cite{materassi09} or bounded nonlinearities~\cite{fazlyab2020safety}.
Moreover, tailored biased QCs for ReLU activation functions were proposed, which include affine terms in the multiplier~\cite{fazlyab2020safety}, copositivity conditions~\cite{Ebihara2024local}, or nonnegativity conditions
enforced on (parts) the multiplier. These usually exploit special properties of ReLU, such as complementarity
or nonnegativity of the output. For completeness, we provide a library of existing biased QCs for general nonlinearities as well as ReLU in Appendix A of this paper.

We now derive the complete set of biased quadratic constraints for ReLU. The new set of QCs extends the (non-incremental) results for ReLU in~\cite{Noori2025} to the biased QC framework.
Denote the matrix~$M_{D}^\text{b}$ as follows:
\begin{equation}\label{eq:MD}
    M_{D}^\text{b} := \bmtx 1 & 0 \\ 0 & D \\ 0 & \frac{1}{2}(I+D)\emtx^\top M^\text{b}\bmtx 1 & 0 \\ 0 & D \\ 0 & \frac{1}{2}(I+D)\emtx,
\end{equation}
where~$D\in \D_{\pm}^m$.
Using the matrix~$M_D^\text{b}$, we can define the complete set~$\MbReLUc$ of
biased QCs for ReLU as follows
\begin{equation}\label{eq:MbReLUc}
    \MbReLUc := \left\{ M^\text{b}\in \Sy^{2m+1}: M_{D}^\text{b}\in \COP^{m+1}\,\forall \, D\in \D_{\pm 1}^m \right\}.
\end{equation}
The complete set of biased QCs for ReLU is characterized by a total of~$2^m$ copositivity conditions. These arise from the possible sign patterns of~$D$. 
\begin{mytheo}\label{thm:complete}
    Consider the repeated nonlinearity~$\Phi: \R^m \to \R^m$ constructed from the ReLU activation function~$\phi:\R \to \R_{\ge0}$. 
    Let $\MbReLUc \subset \Sy^{2m+1}$ be the set defined in \eqref{eq:MbReLUc}.
Then $\Phi$ satisfies the biased QC defined by $M^\text{b}\in \Sy^{2m +1}$ if and only if $M^\text{b}\in\MbReLUc$.
\end{mytheo}
\begin{proof}
    $(\Leftarrow)$ Take any vector~$v\in \R^m$ and let~$w \in \Phi(v)$. Define the vector~$\tilde{v}$ and diagonal matrix~$\tilde{D}$ by
    \begin{equation*}
        \tilde{v}_k = \abs{v_k},\qquad \tilde{D}_{kk} = 
        \begin{cases}
            -1,&\quad v_k <0\\
            1,&\quad v_k\ge 0
        \end{cases}
    \end{equation*}
Then
\begin{equation*}
    v = \tilde{D}\tilde{v},\qquad w = \frac{1}{2}(I + \tilde{D})\tilde{v}
\end{equation*}
and we can write
\begin{equation*}
    \bmtx 1 \\ v\\ w\emtx = {\bmtx 1 & 0 \\ 0 & \tilde{D} \\ 0 & \frac{1}{2}(I +\tilde{D})\emtx}\bmtx 1\\ \tilde{v}\emtx.
\end{equation*}
Thus, by the definition of $M_{\tilde{D}}^\text{b}$ \eqref{eq:MD} we have that 
\begin{equation*}
    \bmtx 1 \\ v \\ w \emtx^\top M^\text{b} \bmtx 1 \\ v \\ w\emtx = \bmtx 1 \\ \tilde{v}\emtx^\top M_{\tilde{D}}^\text{b}\bmtx 1 \\ \tilde{v}\emtx.
\end{equation*}
Since $M^\text{b} \in \MbReLUc$, $M^\text{b}_{\tilde{D}} \in \COP$ for every $\tilde{D} \in \D_{\pm 1}^m$. The vector $[1,\tilde{v}^\top]^\top$ is elementwise nonnegative and therefore
\begin{equation}
    \bmtx 1 \\ v \\ w \emtx^\top M^\text{b} \bmtx 1 \\ v \\ w\emtx = \bmtx 1 \\ \tilde{v}\emtx^\top M_{\tilde{D}}^\text{b}\bmtx 1 \\ \tilde{v}\emtx \ge 0.
\end{equation}
As the choice of $v\in \R^m$ was arbitrary, $\Phi$ satisfies the biased QC defined by $M^\text{b}$.

$(\Rightarrow)$
We prove the reverse direction via contraposition. Suppose ~$M^\text{b}\notin \MbReLUc$. Then there exists a $\tilde{D} \in \D_{\pm 1}^m$ such that $M_{\tilde{D}}^\text{b}\notin \COP^{m+1}$. Hence, there exists a nonzero vector 
\begin{equation*}
x = [x_0, z^\top]^\top 
\end{equation*}
such that
\begin{equation*}
    x^\top M_{\tilde D}^\text{b} x < 0.
\end{equation*}
By continuity, the violating vector can be chosen with $x_0 >0$. Since the quadratic form is homogeneous, we can normalize $x_0 =1$ and write $x = [1, \tilde v^\top]^\top$, where $\tilde v \ge 0$.

Now define the vectors
\begin{equation*}
    v:= \tilde D \tilde v, \quad w:=\frac{1}{2}(I + \tilde D)\tilde v.
\end{equation*}
By construction, $w = \Phi(v)$ since $\tilde v \ge 0$ and $\tilde D \in \D_{\pm 1}^m$. However,
\begin{equation}
    \bmtx 1 \\ v \\ w \emtx^\top M \bmtx 1 \\ v \\ w\emtx = \bmtx 1 \\ \tilde{v}\emtx^\top M_{\tilde{D}}^\text{b}\bmtx 1 \\ \tilde{v}\emtx < 0.
\end{equation}
Thus, $\Phi$ does not satisfy the biased QC defined by $M^\text{b}$ proving the result by contraposition.
\end{proof}


\subsection{Analysis Condition With Extended System}

The proposed framework will bound the Lipschitz constant of the neural network using the stacked system shown in Fig.~\ref{fig:UncSysNew}. 

The condition relies on an affine function $L_e: \Sy^{4m+1}\times \R_{\ge 0} \to \Sy^{2(m+n_u)+1}$ defined as:
\begin{align}
\label{eq:lipLMI}
& L_\text{e}(M^\text{b}, \gamma^2) 
 := 
\left( \bullet \right)^\top
 M^\text{b}
 \bmtx 1 & 0 & 0 \\  G_{21}^\text{s} & G_{22}^\text{s} & G_{23}^\text{s} \\
 0 & 0 & I_{2m}
 \emtx 
\\ \nonumber
& \hspace{0.5in}
+ 
\left( \bullet \right)^\top        
\bmtx 0_{n_y\times 1} & G_{12}^\text{s} & G_{13}^\text{s}\emtx 
\\ \nonumber
& \hspace{0.5in}
-  \gamma^2 \left( \bullet \right)^\top
   \bmtx 0_{n_u\times 1} & T_u & 0_{n_u\times 2m}\emtx, \end{align}
where
\begin{equation*}
    T_u = \bmtx I_{n_u} & -I_{n_u}\emtx.
\end{equation*}
The terms marked $(\bullet)^\top$ in~\eqref{eq:lipLMI} can be inferred from symmetry. The condition with the stacked system is formally stated in Theorem~\ref{thm:elipsdp} below.
\vspacelem
\begin{mytheo}\label{thm:elipsdp}
Consider a well-posed NN $f:\R^{n_u} \to \R^{n_y}$ defined by the LFT $F_\text{l}(G,\Phi)$ in Equation~\ref{eq:LFT}. Assume that the repeated, stacked nonlinearity $\Phi^\text{s}:\R^{2m}\to\R^{2m}$ satisfies the biased QC defined by $M^\text{b}\in \Sy^{4m+1}$. Further assume that there exists a scalar $\gamma \ge 0$ such that $L_\text{e}(M^\text{b},\gamma^2) \preceq 0$. Then the Lipschitz constant of the NN $f$ is bounded by $\gamma$.
\end{mytheo}
\begin{proof}  
Let $\bar{u}$, $\hat{u} \in \R^{m}$ be two inputs to the NN. The NN is assumed to be well-posed, so there exist corresponding solutions $(\bar{v},\bar{w},\bar{y})$ and 
$(\hat{v},\hat{w},\hat{y})$. Multiply $L_\text{e}(M^\text{b},\gamma^2)\preceq 0$ on the left by $[1, \, (u^\text{s})^\top, \, (w^\text{s})^\top]$ and on the right by its transpose. The two-copy LFT equations \eqref{eq:eLFTy} and \eqref{eq:eLFTv} give:
\begin{align}
\label{eq:LipSDIneq}
\bmtx 1 \\ v^\text{s} \\ w^\text{s}\emtx^\top M^\text{b} 
  \bmtx 1\\v^\text{s} \\ w^\text{s}\emtx  
  + \Delta y ^\top \Delta y - \gamma^2 \Delta u^\top \Delta u \le 0.
\end{align}
The first term is non-negative because $\Phi^\text{s}$ satisfies the biased QC defined by $M^\text{b}$.  This yields $\| \Delta y\|_2 \le \gamma \| \Delta u \|_2$, hence the NN Lipschitz constant is upper bounded by $\gamma$.
\end{proof}
\vspacelem
The LMI condition $L_\text{e} \preceq 0$ depends on the bias $G_{21}$ and incorporates nonincremental QCs on the stacked function. We will show, by examples, that this can lead to less conservative (smaller) bounds on the Lipchitz constant than the (incremental) LipSDP LMI condition $L_\text{Lip}\preceq 0$. The reduction in conservatism is problem dependent. However, the next proposition states that the analysis with the stacked system is no more conservative than the analysis with the incremental system.

\vspacelem
\begin{prop}
Consider a well-posed NN $f:\R^{n_u} \to \R^{n_y}$ defined by the LFT $F_\text{l}(G,\Phi)$ in  Equation~\ref{eq:LFT}. Assume that the repeated nonlinearity $\Phi:\R^m\to \R^m$ satisfies the incremental QC defined by $M\in \Sy^{2m}$.  Then the stacked nonlinearity $\Phi^\text{s}:\R^{2m}\to\R^{2m}$ satisfies the biased QC defined by $T_\text{b}^\top M T_\text{b}\in \Sy^{4m+1}$ where 

\begin{align}
T_\text{b} := \bmtx 
0&I_m & -I_m & 0 & 0  \\ 
0&0 & 0 & I_m & -I_m 
\emtx \in \R^{2m \times (4m+1)}.
\end{align}
Moreover,  if $L_\text{Lip}(M,\gamma^2)\preceq 0$ for some scalar $\gamma \ge 0$ then 
$L_\text{e}(T_\text{b}^\top M T_\text{b},\gamma^2)\preceq 0$.
\end{prop}
\begin{proof}  
By assumption, the function $\Phi$ satisfies the incremental QC defined by $M$. This implies, by definition of $T_\text{b}$ that the following inequality holds for all  $v^\text{s}$, $v^\text{s} \in \R^{2m}$:
\begin{align}
0 \le 
\bmtx \Delta v \\ \Delta w \emtx^\top 
M
\bmtx \Delta v \\ \Delta w \emtx
=
\bmtx 1\\v^\text{s} \\ w^\text{s} \emtx^\top
T_\text{b}^\top M T_\text{b}
\bmtx 1 \\ v^\text{s} \\ w^\text{s} \emtx,
\end{align}
Hence the stacked function $\Phi^\text{s}$ satisfies the biased QC defined by 
$T_\text{b}^\top M T_\text{b}$.

Next, define the following $(4m+1)\times(4m+1)$ matrix $U$ that maps $(\Delta u, \Delta w, \hat{u},\hat{w},1)$ to $(1, \bar{u}, \hat{u}, \bar{w}, \hat{w})$:
\begin{align}
U = \bmtx 
0 & 0 & 0 & 0 & 1 \\
I_m & 0 & I_m & 0 & 0 \\
0 & 0 & I_m & 0 & 0 \\
0 & I_m & 0 & I_m & 0 \\
0 & 0 & 0 & I_m & 0 
\emtx
\end{align}
This matrix is a nonsingular, congruence transformation. It follows that
$L_\text{e}(T_\text{b}^\top M T_\text{b},\gamma^2)\preceq 0$
if and only if
$U^\top L_\text{e}(T_\text{b}^\top M T_\text{b},\gamma^2) U
\preceq 0$. Performing this multiplication yields:
\begin{align}
U^\top L_\text{e}(T_\text{b}^\top M T_\text{b},\gamma^2) U
= \bmtx  L_\text{Lip}(M,\gamma^2) & 0 \\ 
        0  & 0_{m+n_u+1} \emtx
\end{align}
Hence  $L_\text{Lip}(M,\gamma^2)\preceq 0$ implies
$L_\text{e}(T_\text{b}^\top M T_\text{b},\gamma^2)\preceq 0$.
\end{proof}
\vspacelem

\section{Computational Approach}\label{sec:comp}

This section provides a computational approach that can be used to find Lipschitz bounds for NNs with repeated ReLU activation functions.

\subsection{General Setup}

Theorem~\ref{thm:elipsdp} provides a sufficient condition to bound the Lipschitz constant of the NN, using the two-copy LFT representation. The Lipschitz bound condition in Theorem~\ref{thm:elipsdp} is an LMI in the variable $\gamma^2$ and any valid parameterization of the biased QC multiplier $M^\text{b}\in\M^\text{b}$. Assume the set of multipliers is described by affine constraints.  In this case, we can find the best (smallest) bound on the Lipschitz constant by solving the following SDP: 
\begin{flalign*}
&\qquad \min_{\substack{M^\text{b},\gamma^2}} \gamma^2&\\
&\qquad \mbox{s.t.}&
\end{flalign*}
\vspace{-25pt}
\begin{subequations}
\label{eq:sdp}
\begin{align}
L_\text{e}(M^\text{b},\gamma^2) \prec 0 \label{eq:Lcond}&\,\\ \gamma^2 \ge 0\\
M^\text{b}\in\mathcal{M}^\text{b}\label{eq:Mcond}&
\end{align}
\end{subequations}

The formulation in~\eqref{eq:sdp} applies to any set of valid biased QCs. Multiplier-specific constraints~\eqref{eq:Mcond} are introduced by the definition of $M^\text{b}$. Larger classes of multipliers generally lead to less conservative bounds at the cost of increased numerical complexity. 

The complete set of biased-QC multipliers $\MbReLUc$ can be exactly characterized for repeated ReLU as given in Theorem~\ref{thm:complete}. This set provides the least restrictive multiplier description for repeated ReLU at the cost of $4^m$ copositivity conditions\footnote{Note that $m$ denotes the dimension of the nonlinear block in a single copy of the original system. The stacked formulation contains two copies and hence has a repeated nonlinearity of dimension $2m$.
Accordingly, the complete characterization contains~$2^{2m}=4^m$ copositivity conditions}.

Verifying that a matrix is copositive is a co-NP-complete problem~\cite{Murty1987}. To derive a tractable SDP we employ the following inner approximation
\begin{equation}
\label{eq:COPdecomp}
   \COP_\text{SPN}^n = \PSD^n + \Sy_{\ge0}^n \subseteq \COP^n.
\end{equation}
In other words, a matrix is copositive if it can be expressed as the sum of a positive semidefinite matrix and an elementwise nonnegative symmetric matrix~\cite{Berman2003}. The relaxation is exact for $n\le 4$. Copositive matrices that admit such a decomposition are referred to as ordinary copositive matrices. Exceptional (non-ordinary) copositive matrices exist for $n\ge 5$, e.g., the Horn matrix (Example 1.30 in \cite{Berman2003}). Thus the inclusion \eqref{eq:COPdecomp} is strict,
in general. Note that the copositive matrices of the complete set $\MbReLUc$ have dimension $2m+1$. Hence the relaxation is guaranteed to be exact for $m=1$.

\subsection{Cutting-plane Algorithm for Complete Set of ReLU}

The complete set of biased QCs for ReLU $\MbReLUc$ encompasses $4^m$ copositivity conditions. We replace these  by the tractable sufficient condition $M_{D_i} \in \COP_\text{SPN}^{2m+1}$. The sufficient condition introduces elementwise nonnegative matrices $N_i \in \Sy^{2m+1}_{\ge 0}$ and adds the following constraints to the SDP:
\begin{equation}
\label{eq:COPdecomp2}
M_{D_i} - N_i \succeq 0\quad \text{and}\quad N_i \ge 0  \mbox{ for } i = 1,\ldots, 4^m.
\end{equation}
These additional $4^m$ PSD constraints render the problem computationally intractable for even moderately sized networks. However,
only a fraction of these constraints are usually active for a given analysis problem. We employ a cutting-plane algorithm to exploit this observation. The algorithm has three main steps. First, solve the main SDP for a working subset of constraints to yield a tentative bound $\gamma^2$ and candidate multiplier $M$. Second, use a separation procedure to search for constraints among the $4^m$ possibilities that are violated by the candidate multiplier $M$. Third, add selected violated constraints, if any, to the master problem. 

The complete cutting-plane algorithm is summarized in Algorithm~\ref{alg:CuttingPlaneShort}.
\begin{algorithm}[t]
\caption{Cutting-plane algorithm}
\label{alg:CuttingPlaneShort}
\small
\begin{algorithmic}[1]

\State Initialize a working set
       $\mathcal{A}_1\subseteq\mathcal{D}_{\pm1}^{2m}$
\State $k\gets1$

\While{true}

    \State Solve the restricted master SDP over $\mathcal{A}_k$
    \State Obtain $M_k$ and $\gamma_k^2$

    \State Evaluate  \Call{Separate}{$M_k,D_i$} for $D_i \in \mathcal{D}_{\pm1}^{2m} \setminus \mathcal{A}_k$.


    \If{no violated constraint is detected}
        \State $M^\star\gets M_k$,
               $\gamma^{\star2}\gets\gamma_k^2$
        \State \textbf{break}
    \EndIf

    \State Rank the violated constraints by separation gap
    \State Select a subset $\mathcal{C}_k$ of violated constraints
    \State $\mathcal{A}_{k+1}
           \gets\mathcal{A}_k\cup\mathcal{C}_k$
    \State $k\gets k+1$

\EndWhile

\end{algorithmic}
\end{algorithm}
Consider the master SDP \eqref{eq:sdp} with 
the complete set of multipliers $\MbReLUc$ for ReLU.  This complete set is defined by $4^m$ copositivity constraints using the set of \emph{all} possible sign patterns  $\D_{\pm 1}^{2m}$. We assume each copositivity constraint is replaced by the inner approximation \eqref{eq:COPdecomp2}. Let $(\gamma^*,M^*)$ denote the Lipschitz bound and optimal multiplier obtained by solving this SDP. 

For the cutting-plane algorithm, we define a working set of constraints for a given iteration $k$ by
\begin{equation*}
    \mcl{A}_k\subseteq \D_{\pm 1}^{2m}
\end{equation*}
The master SDP \eqref{eq:sdp} is solved using a relaxation of the complete set $\MbReLUc$ with constraints \eqref{eq:COPdecomp2} only on the patterns in $\mcl A_k$. This  yields the current multiplier~$M_k$ and the squared bound on the Lipschitz constant $\gamma^2_k$. This relaxation only enforces constraints on $M$ using $\mathcal{A}_k$ and not the full set $\D_{\pm 1}^{2m}$.  This enlarges the set of multipliers and hence the bound at iteration $k$ satisfies $\gamma_k\le \gamma^*$.

The multiplier $M_k$ is then fixed and a separation procedure searches for constraint violations in the neglected set of sign patterns 
$\D_{\pm 1}^{2m} \setminus \mathcal{A}_k$.  For each $D_i \in \D_{\pm1}^{2m}$, define
\begin{equation*}
    S_k(D_i):=L(D_i)^\top M_k L(D_i),
\end{equation*}
with
\begin{equation*}
    L(D_i) = \bmtx1 & 0\\0 & D_i \\ 0 & \frac{1}{2}(I +D_i) \emtx.
\end{equation*}
Feasibility of the corresponding constraint can be checked using the separation SDP
\begin{align*}
    t_k^\star(D_i)
    =
    \min_{t,N}\quad
        & t\\
    \text{s.t.}\quad
        & S_k(D_i)+tI-N\succeq0,\\
        & N\geq0,\\
\end{align*}
The scalar $t_k^\star(D_i)$ quantifies the \emph{separation gap}, i.e., the smallest diagonal shift required for $S_k(D_i)$ to satisfy the
copositivity relaxation:
\begin{equation}
    S_k(D_i) \in \COP_\text{SPN}^{2m+1} \quad\text{iff}\quad t^\star_k(D_i)\le 0.
\end{equation}
The candidate $S_k(D_i)$ is considered infeasible if
\begin{equation*}
    t_k^\star(D_i)>\epsilon_\text{sep},
\end{equation*}
where $\epsilon_{\text{sep}} \ge 0$ is a user-defined numerical tolerance.
The separation gap therefore provides a measure of the violation and can
be used to prioritize violated constraints for addition to the working set.

Solving the SDP for every candidate pattern can itself be costly. We therefore apply a hierarchy of inexpensive sufficient feasibility and infeasibility checks, including elementwise nonnegativity of
$S_k(D_i)$. We only solve the separation SDP for ambiguous candidates. The complete screening procedure is provided in the Appendix B.

For sufficiently small networks, all $4^m$ sign patterns are tested explicitly. Violated constraints are added to the working set and the master SDP is solved again. If no violated pattern is found, the current multiplier satisfies all~$4^m$ constraints and the algorithm terminates.

For larger networks, exhaustive evaluation of all~$4^m$ sign patterns
becomes computationally prohibitive, even with our hierarchy of inexpensive checks. We therefore employ a probabilistic separation procedure combining local Hamming search with global random exploration. The Hamming search is used to efficiently identify promising
violated constraints, while uniformly sampled sign patterns provide global
exploration. The Hamming samples are used only for identification of
cuts and do not contribute to the randomized stopping test.

Let $N_\text{req}$ denote the number of uniformly sampled patterns
used for the randomized stopping test. The required sample size is
selected using a zero-violation binomial bound. 
For prescribed parameters $\varepsilon_\text{v}$ and $\beta_\text{v}$
we choose~$N_\text{req}$ as the smallest integer satisfying
\begin{equation*}
    (1-\varepsilon_v)^{N_\text{req}}\le \beta_v
\end{equation*}
or equivalently
\begin{equation*} N_\text{req} \ge  \frac{\log(\beta_v)} {\log(1-\varepsilon_v)}. 
\end{equation*} 
If the initial Hamming-guided and random separation search detects
no violation, a fresh set of~$N_\text{req}$ sign patterns is sampled
independently and uniformly from~$\D_{\pm 1}^{2m}$. The algorithm  terminates only if this set also contains no violation. Otherwise, selected
violated constraints are added to the working set and the master
SDP is solved again. Details of the adaptive search budget, Hamming-neighborhood construction, and cut-selection procedure are provided
in Appendix C. Note that reaching a prescribed maximum number of 
iterations is not considered convergence.

For this probabilistic method,  the final multiplier $M^\star$ is fixed after termination and the corresponding $S^\star (D_i):= L(D_i)^\top M^\star L(D_i)$ are tested on an independent validation set of size $N_\text{v}\ge N_\text{req}$. 
The minimum choice $N_\text{v} = N_\text{req}$, provides the
prescribed zero-violation guarantee, while a larger validation set
yields a stronger validation result. Define the probability that $M^\star$ violates the imposed copositivity relaxation by
\begin{equation*}
\begin{split}    
    p(M^\star) = \Pr_{D_i\sim 
    \text{unif}(\D_{\pm 1}^{2m})} [S^\star(D_i) \notin \COP_{\text{SPN}}^{2m+1}].
\end{split}
\end{equation*}
If no violation is detected among the $N_\text{v}$ independent validation samples, then $p(M^\star) \le \varepsilon_\text{v}$ with confidence of at least
~$1-\beta_\text{v}$ since $(1-\varepsilon_\text{v})^{N_\text{v}} \le \beta_\text{v}$.

\section{Numerical Examples}\label{sec:numex}
We provide several examples applying the proposed analysis setup. First, we revisit the motivating example from Section~\ref{sec:motivation}. We then apply the approach using different (biased) QCs on a ReLU FNN with a dead neuron and a ReLU FNN with an exponentially growing Lipschitz bound. The section concludes with the analysis of a neural network controller for an inverted pendulum. 

All analyses are performed on a M4 MacBook with $24$ GB memory. All SDPs are solved using a combination of Mosek and YALMIP. All examples apply the following four biased QC parameterizations: 1) the simple affine ReLU QC multiplier denoted by $\MblReLU$ as proposed in~\cite[Lemma 3]{fazlyab2020safety}, 2) the copositive multiplier~$\MblReLUCOP$ proposed in~\cite[Theorem 2]{Ebihara2024local}, 3) the nonnegative multiplier~$\MblReLUNN$ proposed in~\cite[Theorem 2]{Ebihara2024local}, and 4) the complete set multiplier~\eqref{eq:MbReLUc} proposed in Theorem~\ref{thm:complete} of our paper.



\subsection{Motivating Example Revisited}

We first revisit the motivating example from Section~\ref{sec:motivation}. The analysis is now conducted using the stacked system shown in Fig.~\ref{fig:UncSysNew} together with the corresponding analysis condition in Theorem~\ref{thm:elipsdp}. We apply the simple ($\MblReLU$), copositive ($\MblReLUCOP$), nonnegative ($\MblReLUNN$), and complete ($\MbReLUc$) sets of multipliers.

Recall that the simple FNN had a true Lipschiz constant of $4$, $2$, and $2$ for biases of $b^1 =[-1,1]^\top$, $[1,-1]^\top$, and $[0,0]^\top$ (bias-free), respectively. The classical incremental analysis using LipSDP with the complete set of incremental QCs yields a Lipschitz bound of~$4$. This formulation does not account for the bias. It also neglects possible couplings between the incremental pairs $\bar{v}-\hat{v}$ and~$\bar{w}-\hat{w}$ and thus misses certain activation patterns.

We first revisit $b^1 =[-1,1]^\top$. As expected, all four sets of multipliers yield a bound of~$4$ which coincides with the true Lipschitz constant. Next, we consider the bias free case. The simple multiplier yields again a Lipschitz bound of $4$, whereas the nonnegative, copositive, and complete set multipliers all recover the true Lipschitz constant of $2$. Finally, for $[1,-1]^\top$, the simple, nonnegative, and copositive multiplier yield the conservative bound $4$, while the complete set recovers the exact value~$2$. 
Thus, the proposed novel incremental analysis combined with the complete set multiplier recovers the true Lipschitz constant for all three biases.

\subsection{Dead Neuron Example}

A dead neuron in a ReLU NN occurs when a neuron has a non-positive input for all possible inputs, and hence the output remains at zero for all times.
Standard incremental analyses relying on LipSDP fail to recognize dead neurons  (\cite{Kuang2026} or~\cite{gronqvist22}).  We show on a simple example that our analysis approach can calculate tight Lipschitz bounds for NN with dead neurons. Consider the simple MLP defined by
\begin{align*}
& G_{11} = 0, \,\, G_{12} = 0_{2\times 1}, \,\, G_{13} = \bmtx0_{1\times 2}& W^3\emtx, \\
& G_{21} = 0_{4\times 1}, \,\, G_{22} = \bmtx W^1 \\ 0_{2\times 2}\emtx, \,\, G_{23} = \bmtx 0_{2\times 2} & 0\\ W^2 & 0_{2\times 2} \emtx,
\end{align*}
with
\begin{align*}
    & W^1 = \bmtx 1 & 0 \\ 0 & 1\emtx,  \,
    W^2 =\bmtx 1 & 1 \\ -2 & -2\emtx, \, \mbox{ and } 
    W_3 = \bmtx 1 & 10\emtx.
\end{align*}
The example has no bias terms to demonstrate that our analysis is not only beneficial for biased NN, but can also provide tighter bounds as it covers the effects that couple~$(\hat v, \bar v, \hat w, \bar w )$. This NN has a true Lipschitz constant of $L^\star=\sqrt{2}$. This follows because the second neuron in layer two is always inactive, i.e. it is a dead neuron. 

We first perform a incremental analysis with LipSDP using the complete set of incremental QCs for ReLU. Recall that this is the richest class of QC multipliers for ReLU.  Thus this provides the tightest bound on the Lipschitz constant within the LipSDP framework. This analysis yields $\gamma = 28.284$. 

Next, we perform the analysis with the stacked system as proposed in Section~\ref{sec:main}.
We apply the same four multipliers as in the previous example. The simple affine ReLU multiplier yields the same bound as the bias-independent analysis. However, the other three multiplier calculate an upper bound of $\gamma = 1.4142$ and thus recover the exact Lipschitz bound up to the reported numerical accuracy. Using the nonnegative and copositive multiplier requires less than $0.8$ s. Applying the cutting-plane algorithm from Section~\ref{sec:comp} to solve the SDP for the complete set multiplier requires approximately $3.2$ s.
Note that the example has a total of four neurons, i.e., $m=4$. Thus, we verify the cutting-plane algorithm for all possible
$4^m$ activation patterns and the result represents the deterministic upper bound $\gamma$.

\subsection{Exponentially Growing Lipschitz Bound}

The incremental QCs used in the standard Lipschitz bound analysis constitute a smaller class of QCs than the biased QCs used in our proposed incremental analysis. This issue was also reported in \cite[Section 8]{gronqvist22} using a simple bias-free neural network with
\begin{align*}
    G_{11} &= 0, & G_{12} &= 0, & G_{13} &= W^2, \\
    G_{21} &= 0, & G_{22} &= W^1, & G_{23} &= 0
\end{align*}
and weights
\begin{align*}
    W^1 = \bmtx 1 \\ -1\emtx, \quad W^2 = \bmtx 1 & -1\emtx.
\end{align*}

This network has a true Lipschitz constant of $1$, while the standard LipSDP analysis yields a bound of $2$. Moreover, a deep NN built by repeated layers of this network yields an exponentially growing bound of $\gamma_\text{exp} = 2^N$. However, the exact Lipschitz constant of the deep NN remains constant at $1$. This observation is independent of the QC parameterization.

We applied our proposed analysis again with the  
four different QC parameterizations: the simple affine ReLU QC, the copositive multiplier, the nonnegative multiplier, and the complete set multiplier from Theorem~\ref{thm:complete}. The results are reported in Table~\ref{tab:exp}. We observe that the simple affine multiplier yields an exponentially growing Lipschitz bound $\gamma_\text{aff}$. However, the more complex multipliers using nonnegativity or copositivity conditions recover the exact Lipschitz bound of $\gamma_\text{nn}=\gamma_\text{cop}=\gamma_\text{comp}\approx 1$ independent of the depth of the network.
\begin{table}[t]
\centering
\caption{Lipschitz bounds for a network with $N$ repeated hidden layers using different biased QC parameterizations.}
\label{tab:exp}
\begin{tabular}{lrrrr}
\hline
$N$ & $\gamma_\text{aff}$ & $\gamma_\text{nn}$ & $\gamma_\text{cop}$ & $\gamma_\text{comp}$ \\
\hline
$1$ & $2$ & $1$ & $1$ & $1$ \\
$2$ & $4$ & $1$ & $1$ & $1$ \\
$3$ & $8$ & $1$ & $1$ & $1$ \\
$4$ & $16$ & $1$ & $1$ & $1$ \\
\hline
\end{tabular}
\end{table}
Hence, the proposed approach is able to provide accurate solutions for NNs with repeated hidden layers. Note that each layer contains two neurons.  We apply the Hamming-guided-search for the four-layer NN with the complete set of multipliers. The randomized stopping test uses the zero-violation binomial bound with violation probability $\varepsilon_\text{v}=0.01$ and confidence \(1-\beta_\text{v}\), where $\beta_\text{v}=10^{-6}$, i.e., the probability of violating the imposed copositivity relaxation is at most $1\%$ with confidence at least $99.9999\%$.

\subsection{Inverted Pendulum Controller}

We consider the single pendulum benchmark from the AINNCS
category of the 2026 ARCH-COMP competition~\cite{archcomp25,archcomp2026benchmarks}. The provided plant is defined by
\begin{equation}
\dot x_1 = x_2,
\qquad
\dot x_2 = \frac{g}{l}\sin x_1 + \frac{1}{m_\text{p}l^2}\bigl(T-c\,x_2\bigr),
\label{eq:plant}
\end{equation}
with $m_\text{p}=0.5$, $l=0.5$, $c=0$, and $g=1$. The controller is implemented
using a sample-and-hold with sampling period $\Delta t=0.05$\,s. The set of admissible initial
conditions is given by $\mathbb{X}_0=[1.0,1.175]\times[0.0,0.2]$. The safety
specification is
\begin{equation}\label{eq:spec}
\theta(t_k)\in[0,1],\qquad k=10,\dots,20,\qquad t_k=k\Delta t.
\end{equation}
The benchmark provides both continuous and discrete time  specifications. Here, we consider the latter, i.e., the state specification \ref{eq:spec} is enforced on the sampling instants $t_k$.

The benchmark controller $f_\text{T}$ is itself a two-layer ReLU FNN with $25$ hidden neurons per layer, i.e., $50$ hidden neurons in total. For the analysis in this example, we train a smaller ReLU FNN to approximate~$f_\text{T}$. The reduced network has $m=12$ hidden neurons arranged
in two layers of six neurons each. We train the reduced network to imitate the released benchmark controller~$f_\text{T}$ using input-output samples generated from closed-loop rollouts. The reduced
controller achieves an imitation error of $\text{RMSE}=3.2\cdot 10^{-3}$, satisfies~\eqref{eq:spec} on the validation grid, and has an exact Lipschitz constant~$L^\star = 0.8649$, which differs by approximately~$2.5\%$ from that of~$f_\text{T}$. Thus, the reduced controller provides a compact representative of the benchmark controller for comparing the different Lipschitz analyses.  

This reduction has two main motivations.
First, it makes the complete set analysis computationally manageable using the
proposed cutting-plane algorithm in Section~\ref{sec:comp}. Second, the reduced
network allows us to compute the exact global Lipschitz constant as well as the worst-case Lipschitz bound over all biases. We refer to the latter as a "bias-blind" bound. We can assess the tightness of the standard and the proposed analyses against these reference values.

Both reference values are obtained in a combinatorial fashion, i.e., no SDP is
employed. A ReLU FNN is piecewise affine. Thus, each activation pattern
$s\in\{0,1\}^m$ induces a constant Jacobian $J(s)$. The exact constant
$L^\star$ is the largest $\norm{J(s)}_2$ over the patterns that are realizable
by some input. Equivalently,
\begin{equation}
    L^\star = \max_{s\in\mathcal{S}_\text{real}} \norm{J(s)}_2,
\end{equation}
where~$\mathcal{S}_\text{real}$ denotes the set of realizable activation patterns.
Realizability is decided by one linear program per pattern. The
bias-blind value $L_{\text{bb}}$ is the largest $\norm{J(s)}_2$ over all $2^{12}=4096$
patterns. Note that the biases decide which patterns are realizable. However, they never change the Jacobian of a fixed pattern. Moreover, for any given pattern, the
biases can be chosen such that this pattern is realizable, by shifting each
pre-activation to the prescribed sign at some point of the input space. 
Therefore, an analysis that depends only on the weights must remain valid for arbitrary biases cannot exclude any activation pattern. Consequently, $L_\text{bb}$ is
the smallest bound that can be guaranteed using the weights alone.

\begin{table}[t]
\centering
\caption{Lipschitz bounds for the reduced controller with $m=12$.}
\label{tab:results}
\begin{tabular}{llrr}
\hline
& multiplier & bound & time \\
\hline
exact $L^\star$ & enumeration & $0.8649$ & --- \\
bias-blind $L_{\text{bb}}$ & enumeration & $0.9637$ & --- \\
\hline
classical~\eqref{eq:IncSDP} & complete set & $0.9637$ & $<2$\,s \\
\hline
proposed~\eqref{eq:sdp} & affine ReLU QC & $0.9637$ & $<2$\,s \\
& nonnegative & $0.9637$ & $<2$\,s \\
& copositive & $0.9637$ & $<2$\,s \\
& inc. complete set& $\mathbf{0.8649}$ & $300$\,s \\
\hline
\end{tabular}
\end{table}

We first perform the standard LipSDP Lipschitz bound using the incremental system with the reduced controller. The results are summarized in
Table~\ref{tab:results}. This analysis with the
SDP~\eqref{eq:IncSDP} and the complete set multiplier yields an upper bound of
$\gamma_{\text{Lip}}=0.9637$. This value coincides with the bias-blind bound $L_{\text{bb}}$
and is approximately $11.4\%$ higher than the true Lipschitz constant with the
controller. 
Note that this conservatism is not caused by the particular QC parameterization. Any globally valid bias-independent analysis must also be valid for every bias vector associated with the fixed weights and therefore cannot certify a bound below $L_{\mathrm{bb}}$.

Next, we employ the proposed method using the SDP~\eqref{eq:sdp} with the four different QC parameterizations discussed at the beginning of the section: simple ($\MblReLU$), copositive ($\MblReLUCOP$), nonnegative ($\MblReLUNN$), and complete ($\MbReLUc$) sets of multipliers.

The first three multipliers yield an upper bound of
$0.9637$, i.e., the same bound as the bias blind analysis. These analyses require less
than $2$\,s each. In contrast, the biased complete-set multiplier yields $0.8649$, numerically matching the global Lipschitz constant computed independently by activation-pattern enumeration. The cutting-plane algorithm of Section~\ref{sec:comp} requires approximately $300$ s to converge. Since $m=12$, exhaustive separation over $4^{12}$ sign patterns is computationally prohibitive. We therefore apply the Hamming-guided randomized separation procedure of Section~\ref{sec:comp} with~$\varepsilon_\text{v} = 0.01$ and~$\beta_\text{v}=10^{-6}$, which gives~$N_\text{req} = 1375$. The final multiplier passes the randomized stopping test and the subsequent independent validation without detected violations. Hence, the probability of violating the imposed copositivity relaxation is at most~$1\%$ with confidence at least~$99.9999\%$. This example demonstrates that the complete biased QC characterization can eliminate the conservatism caused by neglecting the network biases in global incremental analysis.

\section{Conclusion}
This paper presents a two-copy LFT representation for incremental gain analysis of linear systems interconnected with nonlinearities. The proposed approach explicitly accounts for the effects of bias terms. Moreover, the two-copy formulation enables standard QCs to be applied directly to the stacked nonlinearity rather than restricting the analysis to incremental QCs. Standard QCs admit richer parameterizations, which can further reduce conservatism. We demonstrate the effectiveness of the proposed approach on the Lipschitz analysis of several simple ReLU neural networks. Future work will focus on computing local Lipschitz bounds using the stacked system representation.


\section*{APPENDIX}
\subsection{Examples of Biased-QCs}
We provide several different biased QCs from literature. The first lemma provides a class of biased QCs for bounded nonlinearities, i.e., $\underline{\phi} \le \Phi(v) \le \overline{\phi}$ \cite{fazlyab2020safety}.
\vspacelem
\begin{lem}
    Consider a nonlinearity $\Phi :\R^m \to \R^m$ whose function values are componentwise bounded by $\underline{\phi}$, $\overline{\phi}\in \R^m$ such that $\underline{\phi} \le \Phi(v) \le \overline{\phi}$ for all $v\in\R^m$. Define the following subset of~$\Sy^{2m+1}$
    \begin{equation}
        \Mbbnd(\underline{\phi},\overline{\phi}) :=\left\{ \bmtx -2\underline{\phi}^\top \Lambda\overline{\phi} \!&\! 0 \!&\! (\underline{\phi}+\overline{\phi})^\top\Lambda \\ 0 \!&\! 0 \!&\! 0 \\ \Lambda(\underline{\phi}+\overline{\phi}) \!&\! 0 \!&\! -2\Lambda \emtx : \Lambda\in \D_{\ge 0}^m\right\}.  
    \end{equation}
Then $\Phi$ satisfies the biased QC defined by any~$M^\text{b}\in \Mbbnd(\underline{\phi},\overline{\phi})$.
\end{lem}
\begin{proof}
    The componentwise bounds imply
    \begin{equation*}
        (\Phi(v)-\underline{\phi})^\top \Lambda(\overline{\phi}-\Phi(v))\ge 0
    \end{equation*}
    for any~$\Lambda \in \D_{\geq 0}^m$. Expanding this inequality and multiplying it by two yields the biased QC defined by any $M^\text{b}\in \Mbbnd(\underline{\phi},\overline{\phi})$.
\end{proof}
\vspacelem

Note that this QC follows solely from the componentwise boundedness and does not exploit any additional structure that may arise from the repeated nature of $\Phi$. The next lemma provides a class of biased QCs proposed in \cite{materassi09} for hyperbolic-sector-bounded nonlinearities.

\vspacelem
\begin{lem}\label{lem:hsecbias}
    Consider a repeated nonlinearity $\Phi :\R^m \to \R^m$ constructed from an activation function $\phi:\R \to \R$ satisfying the hyperbolic-sector condition
    \begin{equation}
    (\phi(v) - \alpha v)(\beta v - \phi(v)) +\nu \ge 0 \quad \forall v\in \R,
     \end{equation}
    for some $\alpha,\beta \in \R$ and $\nu\ge 0$.
    Define the following subset of~$\Sy^{2m+1}$
\begin{align}
& \Mbhsec(\alpha,\beta,\nu)
:= \\
\nonumber
& \left\{
\bmtx
2\nu \mathbf{1}^\top\Lambda \mathbf{1} & 0 & 0\\
0 & -2\alpha\beta\,\Lambda & (\alpha+\beta)\,\Lambda\\
0 & (\alpha+\beta)\,\Lambda & -2\Lambda
\emtx \, :  \, \Lambda\in\D_{\ge 0}^m
\right\}.
\end{align}
Then $\Phi$ satisfies the biased QC defined by any~$M^\text{b}\in \Mbhsec (\alpha,\beta, \nu)$
\end{lem}
\begin{proof}
Since each component of the repeated nonlinearity satisfies the
hyperbolic-sector condition,
\begin{equation*}
    (\Phi(v)-\alpha v)^\top
    \Lambda
    (\beta v-\Phi(v))
    +\nu\mathbf{1}^\top(\Lambda)\mathbf{1}\ge0
\end{equation*}
for any $\Lambda\in\D_{\ge0}^m$.
Expanding this inequality and multiplying it by two yields 
\begin{equation*}
    2\nu\mathbf{1}^\top\!\! \Lambda \mathbf{1} - 2\alpha \beta v^\top\!\! \Lambda v +2(\alpha +\beta)v^\top\!\! \Lambda \Phi(v)-2\Phi(v)^\top \Lambda \Phi(v) \ge 0.
\end{equation*}
This corresponds to the biased QC condition \eqref{eq:biasedQC} satisfied by any $M^\text{b}\in\Mbhsec(\alpha,\beta,\nu)$.
\end{proof}
\vspacelem

The biased QC for standard sectors is recovered for $\nu = 0$.

The following lemmas provide common tailored classes of biased QCs for the ReLU activation function, i.e., $\phi(v) =\max(0,v)$ previously proposed in the literature, particularly in the context of local analyses. 
The first lemma provides a simple affine class of biased QCs for ReLU, $\phi(v) = \max(0, v)$ proposed in~\cite{Fazlyab2022}.
\begin{lem}\label{lem:biasrelufaz}
    Consider a repeated nonlinearity~$\Phi: \R^m \to \R^m$ constructed from the ReLU activation function~$\phi: \R \to \R$. Define
\begin{equation}
\begin{split}
&\MblReLU
:=\\
&\Biggl\{
\bmtx
0 & -\nu^\top & (\nu +\eta)^\top\\
-\nu & 0_{m} & \Lambda+ T \\
(\nu +\eta) & \Lambda+ T & -2(\Lambda +T) 
\emtx :\\
&\qquad
\nu,\eta \in\R_{\ge 0}^m,\quad
\Lambda\in\D^m, T\in\DHyp^m
\Biggr\}.
\end{split}
\end{equation}
Then $\Phi$ satisfies the biased QC defined by any~$M^\text{b}\in \MblReLU$
\end{lem}
\begin{proof}
Let $w =\Phi(v)$. The ReLU activation satisfies the componentwise relations
$w\ge 0$, $w-v\ge 0$, and $w_i(v_i -w_i) = 0$, for $i = 1, \ldots,m$.  Since, ReLU is a repeated slope-restricted nonlinearity with the slopes in $[0,1]$ we
have that
\begin{equation*}
    (w_i-w_j)((v_i-v_j) - (w_i-w_j))\ge0, \quad i \neq j.
\end{equation*}
For
\begin{equation}\label{eq:T}
    T = \sum_{1\le i<j\le m} \lambda_{ij}(e_i-e_j)(e_i-e_j)^\top, \lambda_{ij} \ge 0,
\end{equation}
where~$e_i\in \R^m$ is the $i$-th unit vector.
We thus have 
\begin{equation*}
    w^\top T(v-w) = \!\!\!\!\!\!\sum_{1\le i<j\le m}\!\!\!\!\!\lambda_{ij}(w_i-w_j)((v_i-v_j)-(w_i-w_j))\ge 0.
\end{equation*}
Furthermore,
\begin{equation*}
    \nu^\top (w-v) \ge 0, \quad \eta^\top w \ge 0
\end{equation*}
for $\nu, \eta \in \R_{\ge 0}^m$ and 
\begin{equation*}
    w^\top \Lambda(v-w) = 0
\end{equation*}
for any $\Lambda \in \D^m$ due to the ReLU complementarity relation. Consequently, 
\begin{equation*}
    \nu^\top (w-v) + \eta^\top w + w^\top (\Lambda + T)(v-w)\ge 0.
\end{equation*}
Collecting the terms and multiplying by two yields the 
\begin{equation*}
  \bmtx 1 \\ v \\ w\emtx^\top  \bmtx
0 & -\nu^\top & (\nu +\eta)^\top\\
-\nu & 0_{m} & \Lambda+ T \\
(\nu +\eta) & \Lambda+ T & -2(\Lambda +T) 
\emtx \bmtx 1 \\ v \\ w\emtx \ge 0,
\end{equation*}
which proves that $\Phi$ satisfies the biased QC defined by any $M^\text{b} \in \MblReLU$.
\end{proof}

The next lemma provides a class of biased QCs for the ReLU activation
function based on the copositive cone, as proposed in~\cite{Ebihara2024local}.
\begin{lem}\label{lem:ebi_cop}
Consider a repeated nonlinearity $\Phi:\R^m\to\R^m$ constructed from
the ReLU activation function $\phi:\R\to\R$. Define
\begin{equation}\label{eq:Eebi}
E :=
\bmtx
1 & 0_{1\times m} & 0_{1\times m}\\
0_{m\times 1} & -I_m & I_m\\
0_{m\times 1} & 0_m & I_m
\emtx
\end{equation}
and, for $J\in\D^m$,
\begin{equation}\label{eq:Jebi}
\mathcal{J}(J):=
\bmtx
0 & 0_{1\times m} & 0_{1\times m}\\
0_{m\times 1} & 0_m & J\\
0_{m\times 1} & J & 0_m
\emtx.
\end{equation}
Define the following subset of~$\Sy^{2m+1}$:
\begin{equation}
\MblReLUCOP
:=
\left\{
E^\top\left(Q+\mathcal{J}(J)\right)E
:
J\in\D^m,\;
Q\in\COP^{2m+1}
\right\}.
\end{equation}
Then $\Phi$ satisfies the biased QC defined by any
$M^\text{b}\in\MblReLUCOP$.
\end{lem}
\begin{proof}
  Let $w = \Phi(v)$ and define
\begin{equation*}
    z:= E \bmtx 1 \\v \\w\emtx = \bmtx 1 \\ w-v \\ w\emtx.
\end{equation*}
Since $w$ is defined by ReLU we have componentwise
\begin{equation*}
    w - v \ge 0, \quad w\ge 0
\end{equation*}
and thus
\begin{equation*}
    z \ge 0.
\end{equation*}
Hence, for an $Q \in \COP^{2m +1}$
\begin{equation*}
    z^\top Q z \ge 0.
\end{equation*}
Moreover, the ReLU complementarity property
\begin{equation*}
    w_i(w_i-v_i) = 0,\quad i=1,\ldots m,
\end{equation*}
implies for any $J\in \D^m$ that
\begin{equation*}
    (w-v)^\top J w = 0.
\end{equation*}
Therefore,
\begin{equation*}
    z^\top \mathcal{J}(J)z = 2(w-v)^\top J w = 0.
\end{equation*}
It follows that
\begin{equation*}
    z^\top(Q+\mathcal{J}(J))z\ge 0.
\end{equation*}
Substituting the definition of $z$ yields
\begin{equation*}    
\bmtx 1 \\ v \\ w\emtx^\top E^\top (Q+\mathcal{J}(J))E\bmtx 1 \\ v \\ w\emtx \ge 0
\end{equation*}
which proves that $\Phi$ satisfies the biased QC defined by any $M^\text{b} \in \MblReLUCOP$.
\end{proof}

The next lemma provides a tractable class of biased QCs for the ReLU activation function based on elementwise nonnegative multipliers, as proposed in~\cite{Ebihara2024local}.
\begin{lem}\label{lem:ebi_nn}
Consider a repeated nonlinearity $\Phi:\R^m\to\R^m$ constructed from
the ReLU activation function $\phi:\R\to\R$. Define $E$ as in~\eqref{eq:Eebi}
and, for $J\in\D^m$ the matrix~$\mathcal{J}(J)$ as in~\eqref{eq:Jebi}.
Define the following subset of~$\Sy^{2m+1}$:
\begin{equation}
\MblReLUNN
:=
\left\{
E^\top\left(Q+\mathcal{J}(J)\right)E
:
J\in\D^m,\;
Q\in\Sy_{\ge 0}^{2m+1}
\right\}.
\end{equation}
Then $\Phi$ satisfies the biased QC defined by any
$M^\text{b}\in\MblReLUNN$.
\end{lem}
\begin{proof}
Since every elementwise nonnegative symmetric matrix is copositive,
$\Sy_{\ge 0}^{2m+1}\subseteq \COP^{2m+1}$, and therefore
$\MblReLUNN\subseteq\MblReLUCOP$. The result follows directly from
Lemma~\ref{lem:ebi_cop}.
\end{proof}
Hence, $\MblReLUNN$ provides a tractable inner approximation of $\MblReLUCOP$.

\subsection{Efficient Separation Procedure}\label{app:sep}
To avoid solving a semidefinite program for every candidate
$S_k(D_i)$, we introduce a hierarchy of increasingly expensive checks.
First, we check if $S_k(D_i)$ is elementwise nonnegative. In this case,
the candidate is certified feasible by choosing $N_i=S_k(D_i)$, and no
separation SDP is required.

The second check addresses guaranteed violation. Feasibility requires
the diagonal elements of $S_k(D_i)$ to be nonnegative. Otherwise,
there exists no nonnegative matrix $N_i$ which, by subtraction from
$S_k(D_i)$, yields a PSD matrix. This check also provides a lower bound
on the separation gap. Denote the smallest diagonal entry of
$S_k(D_i)$ by
\begin{equation*}
    s_{k,\min}^{\mathrm{diag}}(D_i)
    :=
    \min_j (S_k(D_i))_{jj}.
\end{equation*}
Since
\begin{equation*}
    (S_k(D_i)+tI-N_i)_{jj}\geq0,
    \qquad
    (N_i)_{jj}\geq0,
\end{equation*}
feasibility implies
\begin{equation*}
    t\geq -(S_k(D_i))_{jj}
\end{equation*}
for every $j$, and hence
\begin{equation}
    t_k^\star(D_i)
    \geq
    -s_{k,\min}^{\mathrm{diag}}(D_i).
    \label{eq:diag_lower_bound}
\end{equation}
Thus, a negative diagonal entry directly certifies a violation, and
$-s_{k,\min}^{\mathrm{diag}}(D_i)$ is used as a violation score.

If none of the previous checks is decisive, we check whether
$S_k(D_i)\in\PSD$ within numerical tolerance. In this case,
$N_i=0$ yields
\begin{equation*}
    S_k(D_i)-N_i=S_k(D_i)\succeq0,
\end{equation*}
and the candidate is considered feasible. For numerical efficiency,
positive semidefiniteness is tested by attempting a shifted Cholesky
factorization of
\begin{equation*}
    S_k(D_i)+\epsilon I,
\end{equation*}
where $\epsilon>0$ is a small numerical tolerance.

If the simple PSD check fails, we perform a final inexpensive
feasibility check by constructing a candidate matrix
$N_{\mathrm{triv}}\geq0$. Its diagonal entries are set to zero and its
off-diagonal entries are chosen as the corresponding positive entries
of $S_k(D_i)$. Defining
\begin{equation*}
    P_i:=S_k(D_i)-N_{\mathrm{triv}},
\end{equation*}
the candidate is certified feasible if $P_i\in\PSD$. The same shifted
Cholesky test is used to check positive semidefiniteness within numerical
tolerance.

Candidates that cannot be classified by these inexpensive checks are
evaluated using the separation SDP defined in the main text. A candidate
is classified as violated if
\begin{equation*}
    t_k^\star(D_i)>\epsilon_{\mathrm{sep}}.
\end{equation*}
For these candidates, $t_k^\star(D_i)$ is used as the violation score.

\subsection{Randomized Separation and Cut Selection}
\begin{algorithm}[!t]
\caption{Guided randomized cutting-plane algorithm}
\label{alg:CuttingPlaneLong}
\small
\begin{algorithmic}[1]

\State $\mathcal{A}_1\gets\emptyset$
\State $k\gets1$

\While{true}

    \State Solve the restricted master SDP over $\mathcal{A}_k$
    \State Obtain $M_k$ and $\gamma_k^2$

    \Statex
    \State \textit{Initial guided-plus-random search}

    \StateWrap{Generate Hamming-guided candidates using up to
           the ten most recently added cuts}
    \State Remove duplicates and patterns already in $\mathcal{A}_k$

    \State Retain at most
    $\lfloor 0.25N_{\mathrm{search}}\rfloor$
    guided candidates

    \State Let $N_{\mathrm{guided}}$ be the number of retained
           guided candidates

    \State $N_{\mathrm{rand}}
           \gets
           \max\{0,
           N_{\mathrm{search}}-N_{\mathrm{guided}}\}$

    \State Generate $N_{\mathrm{rand}}$
           uniform-random candidates from
           $\mathcal{D}_{\pm1}^{2m}$

    \StateWrap{Evaluate all guided and random candidates using
           \Call{Separate}{$M_k,D_i$}}

    \StateWrap{Let $r_k$ be the number of violations detected
           in the initial guided-plus-random search}

    \Statex
    \State \textit{Adaptive random exploration}

    \State Determine $N_{\mathrm{extra}}$ according to
    \[
    N_{\mathrm{extra}}
    =
    \begin{cases}
        0,
            & r_k>20,\\
        \min\{50{,}000,N_{\mathrm{req}}\},
            & 10<r_k\leq20,\\
        \min\{100{,}000,N_{\mathrm{req}}\},
            & 3<r_k\leq10,\\
        N_{\mathrm{req}},
            & 0\leq r_k\leq3
    \end{cases}
    \]

    \If{$N_{\mathrm{extra}}>0$}
        \StateWrap{ Draw $N_{\mathrm{extra}}$
               additional uniform-random patterns
               independently of the initial search}
        \State Evaluate them using
               \Call{Separate}{$M_k,D_i$}
    \EndIf

    \State \textit{Randomized stopping test}

    \If{no violation has been detected}

        \StateTwo
            {Draw a fresh stopping set of
             $N_{\mathrm{req}}$ patterns}
            {independently and uniformly from
             $\mathcal{D}_{\pm1}^{2m}$}

        \State Evaluate the stopping set using
               \Call{Separate}{$M_k,D_i$}

        \If{no violation is detected}
            \State $M^\star\gets M_k$
            \State $\gamma^{\star2}\gets\gamma_k^2$
            \State \textbf{break}
        \EndIf

    \EndIf

    \Statex
    \State \textit{Cut selection and working-set update}

    \StateWrap{Retain up to $150$ unique violated patterns
           in the candidate pool}
    \StateWrap{Rank the candidates by the violation score returned by
           \Call{Separate}{$M_k,D_i$}}

    \StateWrap{Select up to $50$ candidates in decreasing
           order of violation score}
    \StateWrap{subject to Hamming distance at least two
               from every cut already selected in this iteration}

    \State Denote the selected cuts by $\mathcal{C}_k$

    \State $\mathcal{A}_{k+1}
           \gets
           \mathcal{A}_k\cup\mathcal{C}_k$

    \State $k\gets k+1$

    \If{$k$ exceeds the prescribed maximum number of iterations}
        \State Terminate without declaring convergence
    \EndIf

\EndWhile
\State \textit{Independent validation}

\State Draw an independent validation set of size
       $N_v\geq N_{\mathrm{req}}$

\State Evaluate $S^\star(D_i)$ for all validation patterns

\end{algorithmic}
\end{algorithm}[h!]
For larger networks, exhaustive evaluation of the $4^m$ sign patterns is
replaced by a combination of Hamming-guided and uniform-random
exploration. The Hamming search is intended to locate additional
violations near constraints that have recently been added to the working
set, whereas the uniform-random samples provide global exploration.

At each iteration, up to the ten most recently added working-set
patterns are used as parent patterns. For each parent, all
Hamming-distance-one neighbors are generated together with up to $100$
randomly selected Hamming-distance-two neighbors. The Hamming-guided
candidates are restricted to at most $25\%$ of the initial search budget
\begin{equation}
    N_{\mathrm{search}}
    =
    \min\{20{,}000,N_{\mathrm{req}}\}.
    \label{eq:app_Nsearch}
\end{equation}
The remaining initial search budget is allocated to patterns sampled
uniformly from $\D_{\pm1}^{2m}$. The Hamming-guided samples are used only
for identifying candidate cuts and are not used in the randomized
stopping test.

Let $r_k$ denote the number of violations detected during this initial
guided-plus-random search. The amount of additional uniform-random
exploration is selected adaptively according to
\begin{equation}
N_{\mathrm{extra}}
=
\begin{cases}
    0,
        & r_k>20,\\[1mm]
    \min\{50{,}000,N_{\mathrm{req}}\},
        & 10<r_k\leq20,\\[1mm]
    \min\{100{,}000,N_{\mathrm{req}}\},
        & 3<r_k\leq10,\\[1mm]
    N_{\mathrm{req}},
        & 0\leq r_k\leq3.
\end{cases}
\label{eq:app_adaptive}
\end{equation}
Thus, broader random exploration is performed as violations become sparse. For $0 \le r_k \le 3$, $N_\text{req}$ additional patterns are sampled independently of the initial
search. These additional samples are used only for identifying
further violated constraints. If no violation is detected after the completed guided-plus-random search, a separate fresh set of $N_\text{req}$ patterns is drawn independently and uniformly for the randomized stopping test, as described in the main text.

A candidate pool of up to $150$ unique violated patterns is retained.
The candidates are ranked according to the violation score returned by
the separation procedure. Up to $50$ constraints are then selected
for addition to the working set. To avoid adding strongly clustered
constraints in the same iteration, a candidate is accepted only if its
Hamming distance from every constraint already selected in that
iteration is at least two. If $\mcl C_k$ denotes the selected set, the
working set is updated according to
\begin{equation*}
    \mcl A_{k+1}
    =
    \mcl A_k\cup\mcl C_k.
\end{equation*}

In the randomized implementation, the working set is initially empty.
Once a violated constraint has been identified, it is retained in the
working set; no pruning is performed. If no violation is detected during
the guided-plus-random search, including any adaptive additional samples, a fresh independent set of
$N_{\mathrm{req}}$ uniformly sampled patterns is evaluated. Convergence
is declared only if no violation is detected in this stopping set.
Reaching the prescribed maximum number of iterations is not considered
convergence.

After convergence, the final multiplier $M^\star$ is held fixed and an
independent validation set of size $N_v\geq N_{\mathrm{req}}$ is
evaluated. These validation samples are independent of all samples used
during the cutting-plane iterations.
The complete cutting-plane procedure including Hamming-guided-search and randomization is summarized in
Algorithm~\ref{alg:CuttingPlaneLong}.


\bibliographystyle{IEEEtran}
\bibliography{ReferencesGWshortArxiv}

\end{document}